\documentclass[]{interact}

\usepackage{epstopdf}
\usepackage[caption=false]{subfig}

\usepackage[numbers,sort&compress]{natbib}

\usepackage{float}

\usepackage{hyperref}

\bibpunct[, ]{[}{]}{,}{n}{,}{,}
\renewcommand\bibfont{\fontsize{10}{12}\selectfont}
\makeatletter
\def\NAT@def@citea{\def\@citea{\NAT@separator}}
\makeatother

\theoremstyle{plain}

\theoremstyle{definition}

\theoremstyle{remark}

\begin{document}

\articletype{ORIGINAL MANUSCRIPT}

\title{A framework of the transport model for high-order eddy viscosity tensor in two-dimensional turbulent flow}

\author{
\name{Xingguang Zhou, Dalin Zhang*, Wenxi Tian, Guanghui Su and Suizheng Qiu\thanks{CONTACT Prof. Dalin Zhang Email: dlzhang@mail.xjtu.edu.cn} }
\affil{State Key Laboratory of Multiphase Flow in Power Engineering, Xi’an Jiaotong University, 710049, Xi’an, Shaanxi, China}
}

\maketitle

\begin{abstract}
Motivated by the concept of eddy viscosity tensor in improved Boussinesq hypothesis, a transport model of high-order eddy viscosity tensor in 2D-3C turbulence structure is derived from the second-order moment model by tensorial analysis. Compared with the previous high-order eddy viscosity tensor mathematical model \cite{Dubrulle1991,Gama1994,Wirth1995} for turbulent flow under special operating conditions, we have obtained a general high-order eddy viscosity tensor transport model, which can show the anisotropy and transport characteristics of Reynolds stress. The transport equation of a high-order eddy viscosity tensor includes the transient term, convection term, production term, diffusion term, positive-definite source term, and dissipation term. In addition, when the components of high-order eddy viscosity tensor are positive or negative, the physical meaning of the source terms is also discussed. Two numerical simulation cases are also interpreted to verify the model. First, we apply the transport model of high-order eddy viscosity to the numerical calculation analysis of two-dimensional straight channel turbulent flow, and the results show that this model can effectively characterize the anisotropy of Reynolds stress. With a combination of analytical and numerical methods, we analyze the flow fields and the components of high-order eddy viscosity tensor to investigate the evolution law of high-order eddy viscosity tensor. The characteristics of the principal components and tensorial invariance of dimensionless Reynolds stress anisotropy tensor of two-dimensional straight channel turbulent flow also have been investigated with the Lumley triangle \cite{Lumley1978,Simonsen2005}. The results show that this model has a good agreement with the DNS calculation results for the characteristics of dimensionless Reynolds stress anisotropy tensor. Then, we make the numerical investigation for a two-dimensional planar asymmetric diffuser and compare the results with the experiments of Obi et.al \cite{Obi1993}. The transport model can resolve the separation bubble. The detachment point, attachment point, and separation bubble length are also compared with the experimental measurements, and the results of LES, are shown in good agreement. Finally, other discussions between 2D-3C turbulence structure and EV-RSM are also introduced, which may supply the further calibration of this model in the future. Such a transport model of high-order eddy viscosity tensor is expected to be useful in the theoretical and numerical analysis of high-order eddy viscosity tensor with improved Boussinesq hypothesis for turbulent flow.
\end{abstract}


\section{Introduction}

An incompressible fluid flow is described by the governing equations, as
\begin{subequations} \label{1}
  \begin{equation}
    \frac{\partial u_i}{\partial x_i} = 0, \label{1a}
  \end{equation}
  \begin{equation}
    \rho \frac{\partial u_i}{\partial t} + \rho u_j \frac{\partial u_i}{\partial x_j} = \frac{\partial \tau_{ij}}{\partial x_j} + f_i.   \label{1b}
  \end{equation}
\end{subequations}
In which $u_i$ is the velocity, $\rho$ is the density of the fluid, $\tau_{ij}$ is a second-order symmetric tensor which represents the forces acting on surfaces, $f_i$ is the external body forces such as gravity. For a Newtonian fluid, Navier and Stokes proposed a series of assumptions about modeling the stress tensor $\tau_{ij}$. The constitutive equation of $\tau_{ij}$ is derived as
\begin{equation}
  \tau_{ij} = -p\delta_{ij} + c_{ijkl}\frac{\partial u_k}{\partial x_l}, \label{2}
\end{equation}
where $p$ is the pressure, $\tau_{ij}^V = c_{ijkl}\frac{\partial u_k}{\partial x_l}$ is the viscous stress tensor, which is a linear homogeneous tensor function of the local velocity gradient. $c_{ijkl}$ is a fourth-order tensor used to characterize the molecular viscosity of the fluid and the index of $i$ and $j$ are symmetric. For general fluids, $c_{ijkl}$ is usually reduced to an isotropic fourth-order tensor, with symmetric index of $i$ and $j$
\begin{equation}
  c_{ijkl} = \nu \delta_{ij}\delta_{kl} + \mu\left(\delta_{ik}\delta_{jl} + \delta_{il}\delta_{jk}\right) \label{3}
\end{equation}
in which $\nu$ and $\mu$ are both scalars, $\delta_{ij}$ is the Kronecker symbol. Substituting Equation \eqref{3} into Equation \eqref{2}, the constitutive relation of incompressible fluid flow can be derived as 
\begin{equation}
  \tau_{ij} = -p\delta_{ij} + 2\mu S_{ij}, \label{4}
\end{equation}
where $\mu$ is the dynamic viscosity, and $S_{ij} = \frac{1}{2}\left(\frac{\partial u_i}{\partial x_j} + \frac{\partial u_j}{\partial x_i}\right)$ is the rate of the strain tensor. Therefore, the Navier-Stokes equation for incompressible fluid flow can be derived as 
\begin{equation}
  \frac{\partial u_i}{\partial t} + u_j \frac{\partial u_i}{\partial x_j} = -\frac{1}{\rho}\frac{\partial p}{\partial x_i} + \nu \frac{\partial}{\partial x_j}\frac{\partial u_i}{\partial x_j} + g_i, \label{5}
\end{equation}
in which $\nu$ is the kinematic viscosity, and $g_i$ is the gravitational acceleration.

For turbulent flow, the Reynolds time-averaging method is usually used to get the RANS (Reynolds-Averaged Navier-Stokes) equation to describe the motion of fluids in large-scale turbulent flow
\begin{equation}
  \frac{\partial \overline{u}_i}{\partial t} + \overline{u}_j \frac{\partial \overline{u}_i}{\partial x_j} = -\frac{1}{\rho}\frac{\partial \overline{p}}{\partial x_i} + \nu \frac{\partial}{\partial x_j}\frac{\partial \overline{u}_i}{\partial x_j} - \frac{\partial \overline{u'_i u'_j}}{\partial x_j} + g_i, \label{6}
\end{equation}
where $\overline{u}_i$ is the time-averaging velocity, $u'_i = u_i - \overline{u}_i$ is the fluctuating velocity.

In RANS, an unclosed term called Reynolds stress $\overline{u'_i u'_j}$ has occurred. Analogous to the constitutive relationship between stress tensor and rate of strain tensor in laminar flow as Equation \eqref{4}, the eddy viscosity model (EVM) is proposed by Boussinesq hypothesis \cite{Boussinesq1877} to model the Reynolds stress
\begin{equation}
  -\overline{u'_i u'_j} = -\frac{2}{3}k\delta_{ij} + \nu_t\left(\frac{\partial \overline{u}_i}{\partial x_j} + \frac{\partial \overline{u}_j}{\partial x_i}\right), \label{7}
\end{equation}
in which $k=\frac{1}{2}\overline{u'_i u'_i}$ is the turbulent kinetic energy, $\nu_t$ is a scalar called the eddy kinematic viscosity. Equation \eqref{7} relates Reynolds stress to time-averaging quantities. According to the dimensional analysis, many two-equations models, such as standard $ k-\varepsilon$ \cite{Launder1974} and standard $k-\omega$ \cite{Wilcox1988}, have been proposed to solve $\nu_t$. However, in these two-equations models, only partly transportation property of $\overline{u'_i u'_j}$ can be considered through $k$ equation and $\varepsilon$ equation or $\omega$ equation \cite{Hanjalic2021}. The anisotropic, relaxation, and historical effect of $\overline{u'_i u'_j}$ are not well characterized. 

For anisotropic of $\overline{u'_i u'_j}$, the NLEVM (Non-Linear Eddy Viscosity Model) \cite{Gatski2000} is proposed. In NLEVM, $\overline{u'_i u'_j}$ is modeled with a series of tensor basis \cite{Lund1992, Pope1975}
\begin{equation}
  \overline{u'_i u'_j} = \frac{2}{3}k\delta_{ij} + \sum_{n=1}^N\alpha'_n T_{ij}^{\left(n\right)}, \label{8}
\end{equation}
where $\alpha'_n$ is the expansion coefficients, and $T_{ij}^{\left(n\right)}\left(n=1,\cdots, N\right)$ is a given tensor basis. According to the definition of dimensionless Reynolds stress anisotropy tensor $b_{ij} = \overline{u'_i u'_j} / \overline{u'_k u'_k} - \frac{1}{3}\delta_{ij}$, Equation \eqref{8} can be simplified to
\begin{equation}
  b_{ij} = \frac{\overline{u'_i u'_j}}{\overline{u'_k u'_k}} = \sum_{n=1}^N\alpha'_n T_{ij}^{\left(n\right)}. \label{9}
\end{equation}
With the assumption of the functional dependency of $b_{ij} = b_{ij}\left(S_{kl}, W_{kl}, k/\varepsilon\right)$, in which $W_{ij} = \frac{1}{2}\left(\frac{\partial u_i}{\partial x_j} - \frac{\partial u_j}{\partial x_i}\right)$ is the vorticity tensor, $T_{ij}^{\left(n\right)}$ can be solved by isotropic tensor function transformation and Cayley-Hamilton theorem. The expansion coefficients $\alpha'_n$ need to be calibrated by DNS data or experimental data \cite{Gatski2000}. NLEVM has been interpreted in many research works, such as Shih quadratic model \cite{Shih1995} and Lien cubic model \cite{Lien1996}. However, for the tensor basis $T_{ij}^{\left(n\right)}$, some assumptions are made and cannot guarantee the choice of the best representation basis \cite{Gatski2000}. Regarding the property of NLEVM, the more terms of tensor series expansion, the more accurate the calculation result is. However, more terms of tensor series expansion will make $\alpha'_n$ more difficult to be calibrated by DNS or experimental measurements, and the oscillation of non-physical phenomena may occur \cite{Liu1996}.

In the aspect of RANS, Reynolds stress is a physical quantity with large-scale behavior that is characterized by the flow field. And the stress tensor in laminar flow is a molecular-scale behavior. Therefore, Reynolds stress does not have the same instantaneous property as the stress tensor in laminar flow. To characterize the anisotropy, historical effect, and relaxation property of Reynolds stress and to avoid the assumptions and higher-order tensor basis expansion in NLEVM. The concept of eddy viscosity tensor \cite{Dubrulle1991,Gama1994, Hanjalic2020, Wirth1995, Wang1995} is proposed. 

To date, many researches on high-order eddy viscosity tensor have been proposed. Dubrulle and Frisch \cite{Dubrulle1991} use the multiscale formalism theory to discuss the eddy viscosity tensor for the incompressible flow of arbitrary dimensionality subject to forcing periodic in space and time, and the explicit expressions of eddy viscosity tensor are given for basic flow with low Reynolds numbers, and when the basic flow is layered. Gama et al. \cite{Gama1994} make the theoretical and numerical research of negative eddy viscosity in isotropic forced two-dimensional flow. But the flow is also assumed that the basic flow in space-time periodic with parity-invariance. Wirth et al. \cite{Wirth1995} give the detailed theoretical and numerical results for eddy viscosity tensor of three-dimensional forced spatially periodic incompressible flow with weak large-scale perturbation theory. Wang and Jiang \cite{Wang1995} consider the high-order eddy viscosity tensor for the modeling of shear turbulence. The high-order eddy viscosity tensor is inversely solved by using Moore-Penrose generalized inverse matrix theory in abstract algebra for Reynolds stress. However, this method is not good for the closure of turbulence model. And the high-order eddy viscosity tensor directly acts on the rate of strain tensor rather than the velocity gradient tensor, which is a simplified form of the high-order eddy viscosity tensor. 

Most of the previous theoretical works are aimed at the simple and basic turbulent flow. Both the high-order eddy viscosity tensor and the eddy viscosity coefficient are the modeling methods to characterize Reynolds stress. Moreover, Reynolds stress is a conserved quantity, which has a conservation equation and is known as the second-order moment transport equation. Hence, we consider that the high-order eddy viscosity tensor can also be regarded as a conserved quantity, which means that there should be a conservation equation for high-order eddy viscosity tensor. This conservation equation should be an explicit fourth-order tensor transport equation and contains the transient term, convection term, generation term, dissipation term, and other source terms to reflect the physical evolution of high-order eddy viscosity tensor. In this article, we carry out the study of this transport equation for high-order eddy viscosity tensor and try to explain the evolution of high-order eddy viscosity in the aspect of physics. 

The rest of this article is organized as follows. In Section 2, we build the constitutive relation between Reynolds stress and velocity gradient with the fourth-order eddy viscosity tensor in 2D-3C turbulence structure. Then we investigate the second-order moment transport equation of Reynolds stress and make the closure by LRR-IP \cite{LRRIP} method. The constitutive relation is regarded as the tensor function. With the mathematical transformation and tensorial analysis of the modeled second-order moment transport equation, the transport equation of high-order eddy viscosity tensor in 2D-3C turbulence structure is obtained, and each term of the transport equation is analyzed in the aspect of physics. In Section 3, numerical analysis and model validation are carried out. First, we study the evolution of high-order eddy viscosity tensor in the two-dimensional straight channel turbulence. The characteristics of the principal components and tensorial invariance of dimensionless Reynolds stress anisotropy tensor are also investigated with the Lumley triangle \cite{Lumley1978,Simonsen2005}. For complex turbulent flow, the two-dimensional turbulent flow in an asymmetric planar diffuser also be calculated, and the results are validated by the experimental measurements. With the transport model of high-order eddy viscosity tensor, the distribution of each independent component of high-order eddy viscosity tensor in an asymmetric planar diffuser is also obtained. In Section 4, we review the assumptions that are made in the constitutive relation between Reynolds stress and velocity gradient with high-order eddy viscosity tensor in 2D-3C turbulence structure. The transport equation in a strict form of eddy viscosity coefficient is derived, and the relation of the Spalart-Allmaras one-equation model \cite{Spalart1994} is also introduced. Finally, we summarize our results and derivations of this framework of the transport model for high-order eddy viscosity tensor. The outlook and potential extensions are also discussed in this study.

\section{Mathematical method and tensorial analysis}
\subsection{Constitutive relation and assumptions}
Analogous to the constitutive of laminar flow as Equation \eqref{2},  $\overline{u'_i u'_j}$ can be modeled by
\begin{equation}
  -\overline{u'_i u'_j} = -\frac{2}{3}k\delta_{ij} + \zeta_{ijkl}\frac{\partial u_l}{\partial x_k}, \label{10}
\end{equation}
in which $\zeta_{ijkl}$ is a fourth-order eddy viscosity tensor.

Considering a statistically two-dimensional flow in which statistics are independent of $x_3$, and which is statistically invariant under reflections of the $x_3$ coordinate axis. The PDF (Probability Density Function) of velocity can be implied as
\begin{subequations}
  \begin{equation}
    \frac{\partial f}{\partial x_3} = 0, \label{11a}
  \end{equation}
  \begin{equation}
    f\left(u_1, u_2, u_3; x_1, x_2, x_3, t\right) = f\left(u_1, u_2, -u_3; x_1, x_2, x_3, t\right). \label{11b}
  \end{equation}
\end{subequations}
At $x_3 = 0$, we can get $\overline{u}_3 = -\overline{u}_3$ and $\overline{u}_3 = 0$. According to the symmetry, we also can get $\overline{u'_1 u'_3} = \overline{u'_2 u'_3} = 0$. Hence for a statistically two-dimensional flow, $\overline{u}_3 = 0$ and Reynolds stress can be derived as
\begin{equation}
  R_{ij} = \overline{u'_i u'_j} = 
  \begin{pmatrix}
    \overline{u'_1 u'_1} & \overline{u'_1 u'_2} & 0 \\
    \overline{u'_2 u'_1} & \overline{u'_2 u'_2} & 0 \\
    0 & 0 & \overline{u'_3 u'_3}
  \end{pmatrix}. \label{12}
\end{equation}
Considering the linear EVM constitutive relation in Equation \eqref{7}, for a two-dimensional turbulent flow, the components of normal Reynolds viscous stress of the $x_3$ coordinate axis can be obtained
\begin{equation}
  R_{33}^V = -\overline{u'_3 u'_3} + \frac{2}{3}k = 2\nu_t \frac{\partial \overline{u}_3}{\partial x_3} = 0. \label{13}
\end{equation}
Therefore, considering the linear EVM constitutive relation, an assumption is made of the transport model of $\zeta_{ijkl}$, that
\begin{equation}
  R_{33}^V = -\overline{u'_3 u'_3} + \frac{2}{3}k = \zeta_{33kl}\frac{\partial \overline{u}_l}{\partial x_k} = 0. \label{14}
\end{equation}
Wang and Jiang \cite{Wang1995} also adopted the same assumption as shown in Equation \eqref{14} when using the simplified high-order eddy viscosity tensor to calculate the two-dimensional turbulent flow. Hence for two-dimensional turbulent flow, the Reynolds viscous stress in the transport model of $\zeta_{ijkl}$ can be obtained
\begin{equation}
  R_{ij}^V = -\overline{u'_i u'_j} +\frac{2}{3}k\delta_{ij} = 
  \begin{pmatrix}
    \zeta_{11kl}\frac{\partial \overline{u}_l}{\partial x_k} & \zeta_{12kl}\frac{\partial \overline{u}_l}{\partial x_k} &  \\
    \zeta_{21kl}\frac{\partial \overline{u}_l}{\partial x_k} & \zeta_{22kl}\frac{\partial \overline{u}_l}{\partial x_k} &  \\
     &  & 0
  \end{pmatrix}. \label{15}
\end{equation}
From Equation \eqref{10}, by contracting the index of $\overline{u'_i u'_j}$, it is obvious that $R_{ij}^V$ is traceless. Therefore, we can get an identity relation
\begin{equation}
  \zeta_{11kl}\frac{\partial \overline{u}_l}{\partial x_k} + \zeta_{22kl}\frac{\partial \overline{u}_l}{\partial x_k} = 0. \label{16}
\end{equation}
Based on the above derivation, $\zeta_{ijkl}$ has 12 independent components because of the symmetric of index $i$ and $j$, that $\zeta_{12kl} = \zeta_{21kl}$.

Therefore, the constitutive relation and mathematical form of high-order eddy viscosity tensor are determined in 2D-3C turbulence structure. Based on the derivation in this section, the transport model of high-order eddy viscosity tensor is studied. And the assumptions which are made for constitutive relation will be also discussed in the following sections.

\subsection{Closure of second-order moment model and derivation of transport equation of $\zeta_{ijkl}$}
From Equation \eqref{10}, only anisotropy of $\overline{u'_i u'_j}$ can be characterized. Spalart and Allmaras \cite{Spalart1994} develop a one-equation model, which builds a transport equation of $\nu_t$ in an empirical way. Spalart-Allmaras one-equation model still uses the linear EVM constitutive relation, which characterizes the isotropic turbulence. However, the transport equation of $\nu_t$ reveals the relaxation and history effect of $\overline{u'_i u'_j}$.

In the following sections, we need to point out the following two points to explain the derivation concisely and clearly:
\begin{enumerate}
  \item If there are no special instructions, time-averaging velocity $\overline{u}_i$ will be simplified to mark as $u_i$, time-averaging pressure $\overline{p}$ will be simplified to mark as $p$. 
  \item The derivation of this article is carried out in the Cartesian coordinate system, which does not need to make the clear specification of contravariant or covariant to the tensor bases and components.
\end{enumerate}

Researching Equation \eqref{10} in the aspects of continuum mechanics and tensorial analysis. From the definition of finite differential of tensor and the derivative of tensor function, since the $\frac{\partial u_l}{\partial x_k}$ is not a symmetric tensor, that Equation \eqref{10} can be written in
\begin{equation}
  \frac{\partial R_{ij}^V}{\partial \left(\frac{\partial u_l}{\partial x_k}\right)} \mathbf{e}_i\mathbf{e}_j\mathbf{e}_k\mathbf{e}_l= \frac{\partial \left(-\overline{u'_i u'_j} + \frac{2}{3} k \delta_{ij}\right)}{\partial \left(\frac{\partial u_l}{\partial x_k}\right)}\mathbf{e}_i\mathbf{e}_j\mathbf{e}_k\mathbf{e}_l=\zeta_{ijkl}\mathbf{e}_i\mathbf{e}_j\mathbf{e}_k\mathbf{e}_l, \label{17}
\end{equation}
in which $\mathbf{e}_i\mathbf{e}_j\mathbf{e}_k\mathbf{e}_l$ is the base of the fourth-order tensor in the Cartesian coordinate system. 

Equation \eqref{17} shows that we regard the constitutive relation of Equation \eqref{10} as the form of tensor function
\begin{equation}
  \mathbf{R}^V = \mathbf{R}^V\left(\nabla \mathbf{u}\right) = \boldsymbol{\zeta} : \nabla \mathbf{u}. \label{18}
\end{equation}
$\mathbf{R}^V$ is the function of $\nabla \mathbf{u}$, where $\nabla \mathbf{u}$ is the argument. In continuum mechanics, such as solid mechanics and the theory of elasticity, the constitutive relation between elastic stress tensor and rate of strain tensor is similar to Equation \eqref{10}. However, the explicit correlation of ordinary material between elastic stress tensor and rate of strain tensor can be measured by experiments \cite{Backus1970,Obermayer2022}. In turbulent fluid flow, the explicit correlation of Reynolds viscous stress field $R_{ij}^V$ and velocity gradient fields $\frac{\partial u_l}{\partial x_k}$ is difficult to measure because of the nonlinear relationship of RANS. For complex turbulent flow, the analytic solution is basically impossible to be obtained. Hence, motivated by Spalart-Allmaras one-equation model to build the transport equation of $\nu_t$. We are also able to build a transport model of high-order eddy viscosity tensor $\zeta_{ijkl}$ to preliminary reveal its evolution law. Different from Spalart-Allmaras one-equation model which is built in an empirical way. We research the second-order moment transport model, to get the transport model of $\zeta_{ijkl}$ in a theoretical way.

The strict form of the second-order moment transport model can be derived from the turbulent fluctuating momentum equation by Reynolds time-averaging
\begin{equation}
  \begin{aligned}
    \frac{\partial \overline{u'_i u'_j}}{\partial t} + u_k \frac{\partial \overline{u'_i u'_j}}{\partial x_k} &= \underbrace{-\left(\overline{u'_i u'_k}\frac{\partial u_j}{\partial x_k} + \overline{u'_j u'_k}\frac{\partial u_i}{\partial x_k}\right)}_{P_{ij}} \\
    &\underbrace{+\overline{\frac{p'}{\rho}\left(\frac{\partial u'_j}{\partial x_i} + \frac{\partial u'_i}{\partial x_j}\right)}}_{\Phi_{ij}} \\
    &\underbrace{-\frac{\partial}{\partial x_k}\left(\overline{u'_i u'_j u'_k}-\nu\frac{\partial \overline{u'_i u'_j}}{\partial x_k} + \frac{\delta_{ik}}{\rho}\overline{u'_j p'} + \frac{\delta_{jk}}{\rho}\overline{u'_i p'}\right)}_{D_{ij}} \\
    &-\underbrace{2\nu\overline{\left(\frac{\partial u'_i}{\partial x_k}\right) \left(\frac{\partial u'_j}{\partial x_k}\right)}}_{\varepsilon_{ij}}, 
  \end{aligned} \label{19}
\end{equation}
where $P_{ij}$ is the production term, $\Phi_{ij}$ is the pressure rate of strain term, $D_{ij}$ is the diffusion term, $\varepsilon_{ij}$ is the dissipation term. By Green-Gauss theorem, Equation \eqref{19} can be written in the integral form with vector notation
\begin{equation}
  \frac{\partial }{\partial t}\int_{V}\mathbf{R}dV + \oint_{A} \mathbf{n}\cdot\mathbf{u}\otimes\mathbf{R}dA = \int_{V}\mathbf{P}dV + \int_{V}\mathbf{\Phi}dV + \oint_{A}\mathbf{n}\cdot\mathbf{D}dA - \int_{V}\boldsymbol{\varepsilon}dV, \label{20}
\end{equation}
where $\mathbf{R} = \overline{u'_i u'_j}\mathbf{e}_i\mathbf{e}_j$ is the Reynolds stress, $\mathbf{n}$ is the unit normal vector of the surface. Substituting Equation \eqref{10} into Equation \eqref{20}, we can get the transport equation of Reynolds viscous stress tensor $\mathbf{R}^V = -\overline{u'_i u'_j}\mathbf{e}_i\mathbf{e}_j + \frac{2}{3}k\delta_{ij}\mathbf{e}_i\mathbf{e}_j$
\begin{equation}
  \begin{aligned}
    \frac{\partial}{\partial t}\int_V\mathbf{R}^VdV - \frac{\partial}{\partial t}\int_V\frac{2}{3}k\mathbf{I}dV &+ \oint_A \mathbf{n}\cdot\mathbf{u}\otimes\mathbf{R}^VdA - \oint_A \mathbf{n}\cdot\mathbf{u}\otimes\frac{2}{3}k\mathbf{I}dA \\
    &=-\int_V\mathbf{P}dV - \int_V\mathbf{\Phi}dV - \oint_A \mathbf{n}\cdot\mathbf{D}dA + \int_V\boldsymbol{\varepsilon}dV,
  \end{aligned} \label{21}
\end{equation}
in which $\mathbf{I} = \delta_{ij}\mathbf{e}_i\mathbf{e}_j$. In Equation \eqref{21}, it is obvious that turbulent kinetic energy transport equation is included. By contracting the index of $i$ and $j$ for Equation \eqref{19}, we can obtain the transport equation of $k$
\begin{equation}
  \frac{\partial k}{\partial t} + u_m\frac{\partial k}{\partial x_m} = \underbrace{-\overline{u'_i u'_m}\frac{\partial u_i}{\partial x_m}}_{P_k} \underbrace{-\frac{\partial }{\partial x_m}\left(\frac{\overline{p'u'_m}}{\rho} + \overline{u'_m u'_i u'_i} - \nu\frac{\partial k}{\partial x_m}\right)}_{D_k} - \underbrace{\nu\overline{\left(\frac{\partial u'_i}{\partial x_m}\right)\left(\frac{\partial u'_i}{\partial x_m}\right)}}_{\varepsilon}, \label{22}
\end{equation}  
in which $P_k$ is the production term, $D_k$ is the diffusion term, $\varepsilon$ is the turbulent kinetic energy dissipation term. Equation \eqref{22} can also be written in the integral form with vector notation
\begin{equation}
  \frac{\partial}{\partial t}\int_VkdV + \oint_A \mathbf{n}\cdot\mathbf{u}kdA = \int_V P_k dV + \oint_A \mathbf{n}\cdot D_k dA - \int_V \varepsilon dV. \label{23}
\end{equation}
Tensor multiplying $\frac{2}{3}\mathbf{I}$ on both sides to Equation \eqref{23}, we can obtain
\begin{equation}
  \frac{\partial }{\partial t}\int_V \frac{2}{3}k\mathbf{I}dV + \oint_A \mathbf{n}\cdot \mathbf{u}\otimes\frac{2}{3}k\mathbf{I}dA = \int_V P_k \otimes\frac{2}{3}\mathbf{I}dV + \oint_A \mathbf{n}\cdot D_k \otimes \frac{2}{3}\mathbf{I}dA - \int_V \varepsilon \otimes\frac{2}{3}\mathbf{I}dV. \label{24}
\end{equation}
Add Equation \eqref{21} and Equation \eqref{24}, the transport equation of Reynolds viscous stress $\mathbf{R}^V$ in strict form is obtained 
\begin{equation}
  \begin{aligned}
    \frac{\partial }{\partial t}\int_V\mathbf{R}^VdV + \oint_A \mathbf{n}\cdot \mathbf{u}\otimes\mathbf{R}^VdA &= \\
    &\underbrace{-\int_V\mathbf{P}dV - \int_V\mathbf{\Phi}dV - \oint_A \mathbf{n}\cdot \mathbf{D}dA + \int_V \boldsymbol{\varepsilon}dV}_{Source \ term \ by \ second-order \ moment \ equation} \\
    &\underbrace{+\int_V P_k \otimes\frac{2}{3}\mathbf{I}dV + \oint_A \mathbf{n}\cdot D_k \otimes \frac{2}{3}\mathbf{I}dA - \int_V \varepsilon \otimes\frac{2}{3}\mathbf{I}dV}_{Source \ term \ by \ turbulent \ kinetic \ energy \ equation}. \label{25}
  \end{aligned}
\end{equation}
Equation \eqref{25} is the strict form that contains unclosed terms. Hence, we need to model these unclosed terms before making the tensorial analysis of $\zeta_{ijkl}$. 

The pressure rate of strain term $\Phi_{ij}$ modeling is crucial and the subject of extensive research \cite{Pope2000}. LRR-IP model \cite{LRRIP} is used in this article, that $\Phi_{ij}$ can be divided into three terms
\begin{equation}
  \Phi_{ij} = \overline{\frac{p'}{\rho}\left(\frac{\partial u'_j}{\partial x_i} + \frac{\partial u'_i}{\partial x_j}\right)} = \Phi_{ij}^r + \Phi_{ij}^s + \Phi_{ij}^w, \label{26}
\end{equation}
where $\Phi_{ij}^r$ is the rapid pressure strain term, $\Phi_{ij}^s$ is the slow pressure strain term, and $\Phi_{ij}^w$ is the wall reflection term. With rapid-distortion theory and Rotta’s model \cite{LRRIP,Pope2000}, the basic model for $\Phi_{ij}$ can be obtained
\begin{equation}
  \Phi_{ij} = \underbrace{-2C_R\varepsilon b_{ij}}_{\Phi_{ij}^r} \underbrace{-C_2\left(P_{ij}-\frac{2}{3}P_k\delta_{ij}\right)}_{\Phi_{ij}^s}, \label{27}
\end{equation}
where $C_R=1.8, C_2 = 0.6$ are empirical constants. The wall reflection term $\Phi_{ij}^w$ is not considered in this basic model. 

For the diffusion tensor of Reynolds stress, Shir model \cite{Shir1973} is adopted in this article
\begin{equation}
  \overline{u'_i u'_j u'_m} = -C_S\frac{k^2}{\varepsilon}\frac{\partial \overline{u'_i u'_j}}{\partial x_m}, \label{28}
\end{equation}
in which $C_S = 0.09$ is an empirical constant. The fluctuating pressure-velocity correlation term $\frac{\delta_{ik}}{\rho}\overline{u'_j p'} + \frac{\delta_{jk}}{\rho}\overline{u'_i p'}$ is ignored \cite{LRRIP}. 

In LRR-IP model, the dissipation tensor of Reynolds stress adopts the homogeneous approximation
\begin{equation}
  \varepsilon_{ij} = 2\nu\overline{\frac{\partial u'_i}{\partial x_k} \frac{\partial u'_j}{\partial x_k}} = \frac{2}{3}\varepsilon \delta_{ij}. \label{29}
\end{equation}
For the diffusion term $D_k$ in turbulent kinetic energy equation, linear eddy diffusion model \cite{Tao2000} is adopted, and ignore the fluctuating pressure-velocity correlation term $\overline{p'u'_m}/\rho$
\begin{equation}
  \overline{u'_m u'_i u'_i} = -C_S\frac{k^2}{\varepsilon}\frac{\partial k}{\partial x_m}. \label{30}
\end{equation}
Substitute Equation \eqref{26} to \eqref{30} into Equation \eqref{25}, the modeled transport equation of Reynolds viscous stress tensor $R_{ij}^V$ can be obtained
\begin{equation}
  \frac{\partial }{\partial t}\int_V R_{ij}^V dV + \oint_A n_m u_m R_{ij}^V dA = P_{ij}^M + D_{ij}^M + \Phi_{R,ij}^M + \Phi_{S,ij}^M+ P_{k,ij}^M + D_{k,ij}^M, \label{31}
\end{equation}
where $P_{ij}^M$, $D_{ij}^M$, $\Phi_{R,ij}^M$, $\Phi_{S,ij}^M$, $P_{k,ij}^M$, and $D_{k,ij}^M$ are the modeled terms of Equation \eqref{25}, which can be shown as 
\begin{subequations}
  \begin{equation}
    P_{ij}^M = -\int_V \left(R_{im}^V\frac{\partial u_j}{\partial x_m} + R_{jm}^V\frac{\partial u_i}{\partial x_m}\right)dV + \int_V\left(\frac{2}{3}k\delta_{im}\frac{\partial u_j}{\partial x_m} + \frac{2}{3}k\delta_{jm}\frac{\partial u_i}{\partial x_m}\right)dV, \label{32a}
  \end{equation}
  \begin{equation}
    D_{ij}^M = \oint_A n_m\left(C_S \frac{k^2}{\varepsilon} + \nu\right)\frac{\partial R_{ij}^V}{\partial x_m} dA - \oint_A n_m\left(C_S \frac{k^2}{\varepsilon} + \nu\right)\frac{\partial }{\partial x_m} \left(\frac{2}{3}k\delta_{ij}\right) dA, \label{32b}
  \end{equation}
  \begin{equation}
    \Phi_{ij}^R = -\int_V C_R \frac{\varepsilon}{k}R_{ij}^VdV, \label{32c}
  \end{equation}
  \begin{equation}
    \Phi_{ij}^S = \int_V C_2\left[\left(R_{im}^V\frac{\partial u_j}{\partial x_m} + R_{jm}^V\frac{\partial u_i}{\partial x_m} - \frac{2}{3}k\delta_{im}\frac{\partial u_j}{\partial x_m} - \frac{2}{3}k\delta_{jm}\frac{\partial u_i}{\partial x_m}\right)-\frac{2}{3}\delta_{ij}R_{rs}^V\frac{\partial u_r}{\partial x_s}\right]dV, \label{32d}
  \end{equation}
  \begin{equation}
    P_{k,ij}^M = \int_V \frac{2}{3} \delta_{ij}R_{rs}^V\frac{\partial u_r}{\partial x_s}dV, \label{32e}
  \end{equation}
  \begin{equation}
    D_{k,ij}^M = \oint_A n_m \frac{2}{3}\delta_{ij}\left(C_k\frac{k^2}{\varepsilon}+\nu\right)\frac{\partial k}{\partial x_m} dA, \label{32f}
  \end{equation}
\end{subequations}
where $C_k = C_S = 0.09$ \cite{LRRIP,Tao2000}. After the modeling of unclosed terms, we can find that the dissipation term does not appear in Equation \eqref{31}. Then, we will make the tensorial derivation of each tensor function term in Equation \eqref{31} with the tensor function relation in Equation \eqref{18}. 

For the transient term in Equation \eqref{31}, taking its derivative to $\nabla \mathbf{u}$, we can get
\begin{equation}
  \frac{d}{d\left(\nabla \mathbf{u}\right)}\frac{\partial }{\partial t}\int_V \mathbf{R}^V dV = \frac{\partial }{\partial t}\int_V\frac{d\mathbf{R}^V}{d\left(\nabla \mathbf{u}\right)}dV = \frac{\partial }{\partial t} \int_V \zeta_{ijkl}dV \mathbf{e}_i \mathbf{e}_j \mathbf{e}_k \mathbf{e}_l. \label{33}
\end{equation}
For the convection term $\oint_A \mathbf{n}\cdot \mathbf{u}\otimes \mathbf{R}^V dA$ in Equation \eqref{31}. Let $\mathbf{H}\left(\nabla \mathbf{u}\right) = \mathbf{n}\cdot \mathbf{u}\otimes \mathbf{R}^V$ and make the tensorial derivative to $\nabla \mathbf{u}$, that we can get
\begin{equation}
  \mathbf{H}'\left(\nabla \mathbf{u}\right):\mathbf{C} = \mathbf{n}\cdot\left(\mathbf{u}':\mathbf{C}\right)\otimes\mathbf{R}^V + \mathbf{n}\cdot\mathbf{u}\otimes\left(\mathbf{R}^{V}\right)':\mathbf{C}, \label{34}
\end{equation}
where $\mathbf{C}$ is an arbitrary second-order tensor from the definition of finite differential of tensor. In Equation \eqref{34}, the unit normal vector of the surface $\mathbf{n}$ is a constant vector function field, therefore $d\mathbf{n}/d\left(\nabla \mathbf{u}\right) = \mathbf{0}$. Since $\mathbf{u} = \mathbf{u}\left(\nabla \mathbf{u}\right)$ and $\mathbf{R}^V = \mathbf{R}^V\left(\nabla \mathbf{u}\right)$ are both the functions of $\nabla \mathbf{u}$, according to the definition of the finite differential of tensor, an implicit term has occurred. However, the transport equation of $\zeta_{ijkl}$ need to be solved explicitly. We can know that $\mathbf{u}' = d\mathbf{u}/d\left(\nabla \mathbf{u}\right)$ is a three-rank tensor with the quotient law of tensor, and the unit of dimension is in meter. It means that the velocity gradient can only work at a certain distance to make the velocity in change. If this distance is very small, we can regard $\mathbf{u}'\left(\nabla \mathbf{u}\right) \approx \mathbf{0}$ through a very dense control volume and mesh. Hence, the convection term after the tensorial derivative can be obtained
\begin{equation}
  \frac{d}{d\left(\nabla \mathbf{u}\right)}\oint_A \mathbf{n}\cdot \mathbf{u}\otimes\mathbf{R}^V dA = \oint_A n_m u_m \frac{\partial R_{ij}^V}{\partial \left(\frac{\partial u_l}{\partial x_k}\right)} dA \mathbf{e}_i \mathbf{e}_j \mathbf{e}_k \mathbf{e}_l = \oint_A n_m u_m \zeta_{ijkl}dA \mathbf{e}_i \mathbf{e}_j \mathbf{e}_k \mathbf{e}_l. \label{35}
\end{equation}
For $P_{ij}^M$ in Equation \eqref{31}, we divide it into two parts
\begin{equation}
  P_{ij}^M = -\underbrace{\int_V\left(R_{im}^V\frac{\partial u_j}{\partial x_m} + R_{jm}^V\frac{\partial u_i}{\partial x_m}\right)dV}_{\mathbf{H}_1} + \underbrace{\int_V \left(\frac{2}{3}k\delta_{im}\frac{\partial u_j}{\partial x_m} + \frac{2}{3}k\delta_{jm}\frac{\partial u_i}{\partial x_m}\right) dV}_{\mathbf{H}_2}. \label{36}
\end{equation}
For $\mathbf{H}_1 = \mathbf{R}^V\cdot \nabla \mathbf{u} + \left(\nabla \mathbf{u}\right)^T \cdot \mathbf{R}^V$, before we make the tensorial derivative of $\nabla \mathbf{u}$, the mathematical transformation needs to be carried out. For two-dimensional incompressible turbulent flow, we can get, 
\begin{subequations}
  \begin{equation}
    H_{1,11} = R_{11}^V\frac{\partial u_1}{\partial x_1} + R_{12}^V\frac{\partial u_1}{\partial x_2} + R_{11}^V\frac{\partial u_1}{\partial x_1} + R_{12}^V\frac{\partial u_1}{\partial x_2} = 2\left(R_{11}^V\frac{\partial u_1}{\partial x_1} + R_{12}^V\frac{\partial u_1}{\partial x_2}\right), \label{37a}
  \end{equation}
  \begin{equation}
    H_{1,12} = R_{11}^V\frac{\partial u_2}{\partial x_1} + \underbrace{R_{12}^V\frac{\partial u_2}{\partial x_2} + R_{21}^V\frac{\partial u_1}{\partial x_1}}_{\nabla \cdot \mathbf{u} = 0} + R_{22}^V\frac{\partial u_1}{\partial x_2} = R_{11}^V\frac{\partial u_2}{\partial x_1} + R_{22}^V\frac{\partial u_1}{\partial x_2}, \label{37b}
  \end{equation}
  \begin{equation}
    H_{1,21} = R_{22}^V\frac{\partial u_1}{\partial x_2} + \underbrace{R_{21}^V\frac{\partial u_1}{\partial x_1} + R_{12}^V\frac{\partial u_2}{\partial x_2}}_{\nabla \cdot \mathbf{u} = 0} + R_{11}^V\frac{\partial u_2}{\partial x_1} = R_{22}^V\frac{\partial u_1}{\partial x_2} + R_{11}^V\frac{\partial u_2}{\partial x_1}, \label{37c}
  \end{equation}
  \begin{equation}
    H_{1,22} = R_{21}^V\frac{\partial u_2}{\partial x_1} + R_{22}^V\frac{\partial u_2}{\partial x_2} + R_{21}^V\frac{\partial u_2}{\partial x_1} + R_{22}^V\frac{\partial u_2}{\partial x_2} = 2\left(R_{21}^V\frac{\partial u_2}{\partial x_1} + R_{22}^V\frac{\partial u_2}{\partial x_2}\right).\label{37d}
  \end{equation} 
\end{subequations}
Equation \eqref{37a} to \eqref{37d} are identity transformations with the continuity equation of incompressible flow and the symmetric property of $\mathbf{R}^V$. For Equation \eqref{37b} and \eqref{37c}, consider $\mathbf{R}^V$ is traceless, which is interpreted in Equation \eqref{16}. Hence Equation \eqref{37b} and \eqref{37c} can be rewritten in
\begin{subequations}
  \begin{equation}
    H_{1,12} = R_{11}^V\frac{\partial u_2}{\partial x_1} + R_{22}^V\frac{\partial u_1}{\partial x_2} = R_{11}^V\left(\frac{\partial u_2}{\partial x_1} - \frac{\partial u_1}{\partial x_2}\right), \label{38a}
  \end{equation}
  \begin{equation}
    H_{1,21} = R_{22}^V\frac{\partial u_1}{\partial x_2} + R_{11}^V\frac{\partial u_2}{\partial x_1} = R_{22}^V\left(\frac{\partial u_1}{\partial x_2} - \frac{\partial u_2}{\partial x_1}\right). \label{38b}
  \end{equation}
\end{subequations}
Therefore, by re-summarizing $\mathbf{H}_1 = \mathbf{R}^V\cdot \nabla \mathbf{u} + \left(\nabla \mathbf{u}\right)^T \cdot \mathbf{R}^V$, we can get
\begin{equation}
  H_{1,ij} = R_{im}^V\frac{\partial u_j}{\partial x_m} + R_{jm}^V\frac{\partial u_i}{\partial x_m} = 2\delta_{ij}R_{\underline{i}m}^V\frac{\partial u_{\underline{i}}}{\partial x_m} + \epsilon_{ij}\omega_Z R_{\underline{i}\underline{i}}^V, \label{39}
\end{equation}
in which $\epsilon_{ij}$ is a two-dimensional Eddington tensor, that $\epsilon_{11} = \epsilon_{22} = 0$, $\epsilon_{12} = 1$, and $\epsilon_{21} = -1$. $\omega_Z = \frac{\partial u_2}{\partial x_1} - \frac{\partial u_1}{\partial x_2}$ is the vorticity of the $x_3$ coordinate axis. It should be noted that, in Equation \eqref{39}, a horizontal line has been added below the index $i$. This means that index $i$ is neither free nor dummy, only plays the following role, and the relevant definitions can be referred to tensor algebra and analysis \cite{Bishop180,Huang2003,Sokolnikoff1964}. 

After the above identical mathematical transformation, the tensorial derivative of $\mathbf{H}_1$ to $\nabla \mathbf{u}$ can be expressed uniquely and explicitly, as
\begin{equation}
  \mathbf{H}'_1 = \int_V\left[\left(2\delta_{ij}R_{\underline{i}m}^V\delta_{\underline{i}l}\delta_{mk} + 2\delta_{ij}\zeta_{\underline{i}mkl}\frac{\partial u_{\underline{i}}}{\partial x_m}\right) + \left(\epsilon_{ij}R_{\underline{i}\underline{i}}^V\Omega_{kl} + \epsilon_{ij}\omega_Z\zeta_{\underline{i}\underline{i}kl}\right)\right]dV \mathbf{e}_i \mathbf{e}_j \mathbf{e}_k \mathbf{e}_l, \label{40}
\end{equation}
where $\mathbf{\Omega}$ is a second-rank tensor, expressed as 
\begin{equation}
  \omega_Z = \mathbf{\Omega}:\nabla \mathbf{u}, \nabla \mathbf{u} = 
  \begin{pmatrix}  
    \frac{\partial u_1}{\partial x_1} & \frac{\partial u_2}{\partial x_1} \\ 
    \frac{\partial u_1}{\partial x_2} & \frac{\partial u_2}{\partial x_2}
  \end{pmatrix} \Rightarrow \frac{d\omega_Z}{d\left(\nabla\mathbf{u}\right)} = \mathbf{\Omega} = \begin{pmatrix}  
    0 & 1 \\ 
    -1 & 0
  \end{pmatrix}. \label{41}
\end{equation}
For the $\mathbf{H}_2$ part in Equation \eqref{36}, we can find that the turbulent kinetic energy $k$ appears. Since turbulent kinetic energy $k$ and turbulent kinetic dissipation rate $\varepsilon$ are not explicit tensor functions of $\nabla \mathbf{u}$. We can consider such a condition, for the steady turbulent kinetic energy transport equation of incompressible fluid flow
\begin{equation}
  \nabla\cdot\left(\mathbf{u}k\right) = P_k + \nabla\cdot\left(\Gamma_k \nabla k\right) - \varepsilon, \label{42}
\end{equation}
where $\Gamma_k = \nu + \nu_t^k$ is considered as the general diffusion coefficient of $k$ equation by linear eddy diffusion model. Add a small disturbance which is induced by velocity gradient to Equation \eqref{42}, we can obtain
\begin{equation}
  \nabla\cdot\left[\left(\mathbf{u}+\delta\mathbf{u}\right)\left(k+\delta k\right)\right] = \left(P_k + \delta P_k\right) + \nabla\cdot\left[\Gamma_k\nabla \left(k + \delta k\right)\right] - \left(\varepsilon + \delta \varepsilon\right), \label{43}
\end{equation}
in which according to the definition of the finite differential of tensor. We can get $\delta \phi = \frac{d\phi}{d\left(\nabla\mathbf{u}\right)} : d\left(\nabla \mathbf{u}\right)$, in which $\phi = \left[\mathbf{u}, k, P_k, \varepsilon\right]^T$. Substitute Equation \eqref{42} into Equation \eqref{43}. Then, volume integration of control volume is performed according to Equation \eqref{43}, and omitting the second-order small quantities, we can get
\begin{equation}
  \oint_A \mathbf{n}\cdot\left[\left(\mathbf{u}\delta k\right)-\left(\Gamma_k\nabla \delta k\right)\right]dA = \int_V \delta P_k dV - \int_V \delta \varepsilon dV. \label{44}
\end{equation}
An assumption is made here for the sake of the simplicity of the model, that
\begin{equation}
  \int_V \delta P_k dV \approx \int_V \delta \varepsilon dV. \label{45}
\end{equation}
This assumption shows that the net flux of $\delta k$, which is induced by the velocity gradient $\nabla \mathbf{u}$, is nearly zero. It means that no matter how the $\nabla \mathbf{u}$ changes, $\delta k$ changes little. Comparing with $\delta \mathbf{R}^V = \frac{d \mathbf{R}^V}{d\left(\nabla \mathbf{u}\right)}:d\left(\nabla \mathbf{u}\right)$, initial fields and boundary conditions are both implemented with $\delta k \approx 0$ and $\delta \mathbf{R}^V \approx \mathbf{0}$. With the same $d\left(\nabla \mathbf{u}\right)$, the change of $\delta k$ is smaller than that of $\delta \mathbf{R}^V$. Therefore, as a zero-order approximation, $dk/d\left(\nabla \mathbf{u}\right)$ will be ignored.

In fact, in the original Boussinesq hypothesis \cite{Boussinesq1877}, the constitutive relation should be written in
\begin{equation}
  -\overline{u'_i u'_j} = -\frac{1}{\rho}p_t\delta_{ij} + \zeta_{ijkl}\frac{\partial u_l}{\partial x_k}, \label{46}
\end{equation}
where $p_t$ is the pressure that is induced by fluctuating velocity, and it is corresponding to the pressure $p$ in Equation \eqref{2}. Then, $p_t = \frac{2}{3}\rho k$ is adopted to the following EVM methods \cite{Hanjalic2021,Pope1975,Tao2000}. Physically, the effect of velocity gradient on $p_t = \frac{2}{3}\rho k$ should be similar to that on $p$. According to the Helmholtz velocity decomposing theorem, the velocity gradient $\nabla \mathbf{u}$ can be decomposed into the rate of strain tensor $\mathbf{S} = \frac{1}{2}\left(\nabla\mathbf{u} + \nabla\mathbf{u}^T\right)$ for elastic deformation and the rotation tensor $\mathbf{W} = \frac{1}{2}\left(\nabla\mathbf{u} - \nabla\mathbf{u}^T\right)$ for rigid rotation. There is no direct and explicit relation between the isotropic pressure and the velocity gradient. Therefore, in the boundary layer theory, the pressure in the fully developed region of turbulence will directly act on the near-wall region under the first-order approximation \cite{Schlichting1996}. The assumption in Equation \eqref{45} is derived from this physical consideration. For turbulent kinetic energy dissipation rate $\varepsilon$, which is only directly acting on $k$ or $p_t$, the same assumption and consideration are adopted. 

After the above derivation, the tensorial derivative of $P_{ij}^M$ to $\nabla \mathbf{u}$ can be obtained
\begin{equation}
  \begin{aligned}
    {\mathbf{P}^M}' = &-\int_V\left[\left(2\delta_{ij}R_{\underline{i}m}^V\delta_{\underline{i}l}\delta_{mk} + 2\delta_{ij}\zeta_{\underline{i}mkl}\frac{\partial u_{\underline{i}}}{\partial x_m}\right) + \left(\epsilon_{ij}R_{\underline{i}\underline{i}}^V\Omega_{kl} + \epsilon_{ij}\omega_Z\zeta_{\underline{i}\underline{i}kl}\right)\right]dV \\
    &+\int_V \frac{2}{3}k\left(\delta_{ik}\delta_{jl} + \delta_{il}\delta_{jk}\right)dV. \label{47}
  \end{aligned}
\end{equation}
From Equation \eqref{31}, it is clearly seen that the tensorial derivative of the remaining terms to $\nabla \mathbf{u}$ can be referred to the procedure and operation in Equation \eqref{47}. Therefore, the tensorial derivative of the remaining terms can refer in Appendix A.

Based on the transport equation of $\mathbf{R}^V$ as Equation (47), and tensorial derivative to $\nabla \mathbf{u}$ as Equation \eqref{33} to \eqref{47} and Appendix A. A transport equation of $\zeta_{ijkl}$ is obtained
\begin{equation}
  \frac{\partial \zeta_{ijkl}}{\partial t} + u_m \frac{\partial \zeta_{ijkl}}{\partial x_m} = P_{ijkl}^{Strain} + P_{ijkl}^{Vortivity} + D_{ijkl}^{\zeta} + S_{ijkl}^{Positive} + E_{ijkl}^{Dissipation}, \label{48}
\end{equation}
in which, 
\begin{subequations}
  \begin{equation}
    P_{ijkl}^{Strain} = \left(C_2-1\right)\left(2\delta_{ij}R_{\underline{i}k}^V\delta_{\underline{i}l} + 2\delta_{ij}\zeta_{\underline{i}mkl}\frac{\partial u_{\underline{i}}}{\partial x_m}\right) + \frac{2}{3}\left(1-C_2\right)\delta_{ij}\left(\zeta_{rskl}\frac{\partial u_r}{\partial x_s} + R_{kl}^V\right), \label{49a}
  \end{equation}
  \begin{equation}
    P_{ijkl}^{Vorticity} = \left(C_2 - 1\right)\left(\epsilon_{ij}R_{\underline{i}\underline{i}}^V\Omega_{kl} + \epsilon_{ij}\omega_Z\zeta_{\underline{i}\underline{i}kl}\right), \label{49b}
  \end{equation}
  \begin{equation}
    D_{ijkl}^{\zeta} = \frac{\partial }{\partial x_m}\left[\left(C_S\frac{k^2}{\varepsilon} + \nu\right) \frac{\partial \zeta_{ijkl}}{\partial x_m}\right], \label{49c}
  \end{equation}
  \begin{equation}
    S_{ijkl}^{Positive} = \frac{2}{3}\left(1-C_2\right)k\left(\delta_{ik}\delta_{jl} + \delta_{il}\delta_{jk}\right), \label{49d}
  \end{equation}
  \begin{equation}
    E_{ijkl}^{Dissipation} = -C_R \frac{\varepsilon}{k}\zeta_{ijkl}. \label{49e}
  \end{equation}
\end{subequations}
The tensor base of Equation \eqref{48} is $\mathbf{e}_i \mathbf{e}_j \mathbf{e}_k \mathbf{e}_l$, and the dimension is $m^2/s^2$. 

The transport model of $\zeta_{ijkl}$ contains five source terms. $P_{ijkl}^{Strain}$ is the strain production term, which indicates that $\zeta_{ijkl}$ is generated by the normal elastic deformation. $P_{ijkl}^{Vorticity}$ is the vorticity production term, that $\zeta_{ijkl}$ is generated by the change of time-averaging vorticity, which is used to generate small-scale turbulence. $D_{ijkl}^{\zeta}$ is the diffusion term, which contains molecular diffusion and turbulence-induced diffusion. $\zeta_{ijkl}$ diffuses from the high turbulent intensity region to the low turbulent intensity region, which makes the turbulence isotropic. Source terms contain a positive-definite source $S_{ijkl}^{Positive}$ and a dissipation term $E_{ijkl}^{Dissipation}$. From the research of Gama et al. and Wirth et al.\cite{Gama1994,Wirth1995}, we can know that $\zeta_{ijkl}$ can be positive or negative. When $\zeta_{ijkl}$ is positive, from constitutive Equation \eqref{10}, it is obvious that the direction of the component of velocity gradient $\frac{\partial u_l}{\partial x_k}$ is opposite to Reynolds stress, which means the momentum transportation by fluctuating velocity is from the high region to the low region.  In this situation, $S_{ijkl}^{Positive}$ performs as a source, and $E_{ijkl}^{Dissipation} < 0$ performs as a dissipation. Therefore, for positive $\zeta_{ijkl}$, flow fields with high turbulent kinetic energy $k$ will increase $\zeta_{ijkl}$. At the same time, flow fields with high turbulent specific kinetic energy dissipation rate $\omega \varpropto \varepsilon / k$ will decrease $\zeta_{ijkl}$. When $\zeta_{ijkl}$ is negative, from constitutive Equation \eqref{10}, we can know that the direction of the component of velocity gradient $\frac{\partial u_l}{\partial x_k}$ is the same to $\mathbf{R}^V$, which means that the momentum transportation by fluctuating velocity is from the low region to the high region. This is not obeying Fick’s law \cite{Davis1995}, which is called by CGT (Converse Gradient Transport) phenomenon \cite{Dixit2021,Jiang2000,Qiu2004}. CGT is first observed by experiments in fluid mechanics \cite{Eskinazi1954} and is characterized by the negative turbulent kinetic energy generation term. It is obvious that linear EVM constitutive relation cannot characterize the CGT, because the eddy viscosity $\nu_t \varpropto k^2 / \varepsilon$ is always positive. When $\zeta_{ijkl}$ is negative, $S_{ijkl}^{Positive}$ is still positive, and acts as a sink; $E_{ijkl}^{Dissipation} > 0$ also acts as a dissipation to prevent the $\zeta_{ijkl}$ more negative. This means the CGT of momentum transportation by fluctuating velocity, which is caused by $\zeta_{ijkl} < 0$, will attenuate faster than the general transport phenomenon. Equation \eqref{48} shows a preliminary transport model of high-order eddy viscosity tensor $\zeta_{ijkl}$ from modeled second-order moment transport equation with tensorial analysis.

From the perspective of properties of Reynolds stress, Equation \eqref{48} can consider the anisotropy of $\overline{u'_i u'_j}$ which compares to Spalart-Allmaras one-equation model. Also, this model can consider the relaxation and history effect of $\overline{u'_i u'_j}$ which compares to non-linear EVM. In the following sections, we will make the numerical analysis and model validation to investigate the characteristics of Equation \eqref{48} in the complex two-dimensional turbulent flow.

\section{Numerical analysis and model validation}
\subsection{Numerical methodology and approximation}
As discussed in the previous sections, Equation \eqref{48} is the high-order eddy viscosity tensor transport model to describe the two-dimensional turbulent flow. Since the model cannot compute the Reynolds stress directly like the second-order moment model. Therefore, $k$ equation and $\varepsilon$ equation still need to be supplemented to close the turbulence model. Contracting the index $i$ and $j$ of the second-order moment transport model will get the transport equation of turbulent kinetic energy. In Section 2.2, Equation \eqref{30} is modeled with the diffusion term of $k$ equation. Hence the modeled $k$ equation will be re-written here
\begin{equation}
  \frac{\partial k}{\partial t} + u_j \frac{\partial k}{\partial x_j} = -\overline{u'_i u'_j}\frac{\partial u_i}{\partial x_j} + \frac{\partial }{\partial x_j}\left[\left(C_k \frac{k^2}{\varepsilon} + \nu\right)\frac{\partial k}{\partial x_j}\right] - \varepsilon. \label{50}
\end{equation}
For ordinary fluid (Prandtl number near 1), $\varepsilon$ equation can be modeled \cite{Chen1998,LRRIP} as
\begin{equation}
  \frac{\partial \varepsilon}{\partial t} + u_j \frac{\partial \varepsilon}{\partial x_j} = -C_{\varepsilon 1}\frac{\varepsilon}{k}\overline{u'_i u'_j}\frac{\partial u_i}{\partial x_j} + \frac{\partial }{\partial x_j}\left[\left(C_\varepsilon \frac{k^2}{\varepsilon} + \nu\right)\frac{\partial \varepsilon}{\partial x_j}\right] - C_{\varepsilon 2}\frac{\varepsilon ^ 2}{k}, \label{51}
\end{equation}
in which $C_\varepsilon = 0.07, C_{\varepsilon 1} = 1.44$, and $C_{\varepsilon 2} = 1.92$ are empirical constants. Thus the turbulence model is closed and can be solved by Equations \eqref{6}, \eqref{10}, \eqref{48}, \eqref{50}, and \eqref{51}. In this article, a turbulence model for two-dimensional turbulent flow in 2D-3C turbulence structure which named as $k-\varepsilon-\zeta$ is developed. 

The finite volume method (FVM) is adopted in this article, and the model is implemented in OpenFOAM. OpenFOAM adopts the SIMPLE (Semi-Implicit Method for Pressure Linked Equations) algorithm \cite{Patankar} to solve the coupling of pressure and velocity during the calculation process. The SIMPLE algorithm solves the inner and outer iterations, the inner iteration solves the large sparse matrix, and the outer iteration updates the coefficients of the equation with the results of the inner iteration. The SIMPLE algorithm does not consider the influence of the adjacent grid velocity correction on the current grid velocity in the discretization. Hence, the field relaxation factors need to be introduced to ensure the stability of the iterative process. However, $k-\varepsilon-\zeta$ model, which is developed in this article, needs to solve 17 equations for two-dimensional turbulent flow in OpenFOAM. The variants contain $u_1$, $u_2$, $p$, $k$, $\varepsilon$, and 12 independent components of $\zeta_{ijkl}$. The second-order moment model directly performs the transport equations for Reynolds stress, which can characterize all properties of Reynolds stress, but its numerical convergence and computational cost have been criticized \cite{Hanjalic2021,Pope1975}. Nonetheless, the second-order moment model only needs to solve 8 equations in a two-dimensional turbulent flow. The variants contain $u_1$, $u_2$, $p$, $\varepsilon$, $\overline{u'_1 u'_1}$, $\overline{u'_1 u'_2}$, $\overline{u'_2 u'_2}$, $\overline{u'_3 u'_3}$. It is obvious that $k-\varepsilon-\zeta$ model even needs to solve 9 more equations than the second-order moment model. Therefore, how to ensure convergence and stability is an urgent problem when solving this large group of strongly coupled nonlinear equations of $k-\varepsilon-\zeta$ model. Through the numerical experiments on OpenFOAM, we find that only adjusting the field relaxation factor and matrix relaxation factor is hard to guarantee the stability of the solution. Hence an approximation is made for the production term in turbulent kinetic energy transport equation
\begin{subequations}
  \begin{equation}
    P_k = -\overline{u'_i u'_j} \frac{\partial u_i}{\partial x_j} = \left(\zeta_{ijkl}\frac{\partial u_l}{\partial x_k}\right)\frac{\partial u_i}{\partial x_j}, \label{52a}
  \end{equation}
  \begin{equation}
    P_k^{apx} = -\overline{u'_i u'_j} \frac{\partial u_i}{\partial x_j} = \nu_t \left(\frac{\partial u_i}{\partial x_j} + \frac{\partial u_j}{\partial x_i}\right)\frac{\partial u_i}{\partial x_j}. \label{52b}
  \end{equation}
\end{subequations}
$\nu_t = C_\mu \frac{k^2}{\varepsilon}$ is the eddy viscosity which is calculated in the form of linear EVM. Equation \eqref{52a} is the strict form of $k-\varepsilon-\zeta$ model in this article, and Equation \eqref{52b} is the approximation form used in numerical simulation with OpenFAOM to ensure the stability and convergence of the numerical solution. It is obvious that $\nu_t > 0$ in $P_k^{apx}$ which cannot characterize the CGT phenomena. From the analysis in Section 2.2, we can know that $\zeta_{ijkl}$ can be positive or negative. Hence the strict form $P_k$ can reveal the CGT phenomena. For the preliminary research of this model, however, $P_k^{apx}$ will be adopted. When the goal is to reveal the CGT phenomenon with this model, a more robust numerical algorithm and the strategy of solution will be adopted to $P_k$, which is not the focus of this article.

\subsection{Numerical simulation and analyzation}
The first case is a two-dimensional straight channel turbulence, and Re is 40000. We calculate the turbulent flow under fully-developed condition with $k-\varepsilon-\zeta$ model and standard $k-\varepsilon$ model \cite{Launder1974}, For two-dimensional channel turbulence, we focus on the properties of its dimensionless Reynolds stress characteristic. 

\begin{figure}[H]
  \centering
  \includegraphics[width=0.8\linewidth]{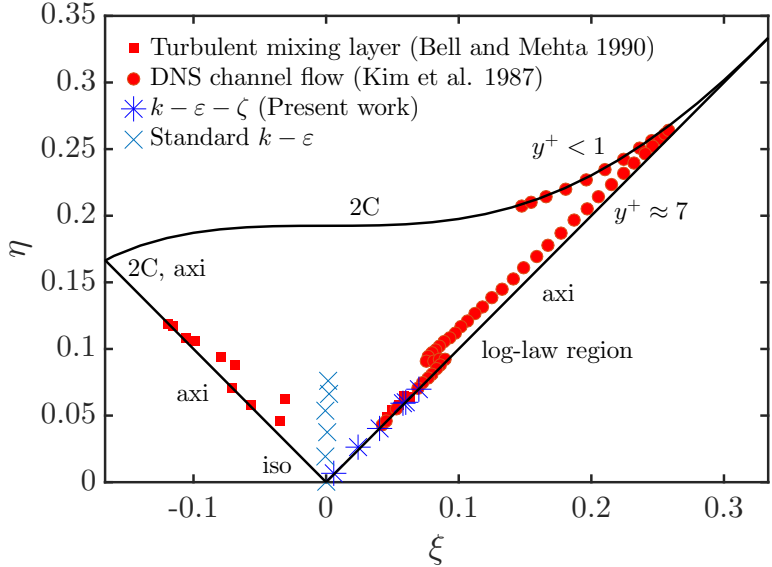}
  \caption{Characteristic quantities distribution of $k-\varepsilon-\zeta$ model and standard $k-\varepsilon$ model in Lumley triangle. Red solid squares represent the turbulent mixing layer obtained by Bell and Mehta \cite{Bell1990}. Red solid dots represent the DNS channel flow obtained by Kim et al. \cite{Kim1987}. Dark blue star-type nodes represent the results of $k-\varepsilon-\zeta$ model. Light blue fork-type nodes represent the results of standard $k-\varepsilon$ model.}
  \label{figure 1}
\end{figure}

Lumley \cite{Lumley1978} investigated the eigenvalues and characteristics of dimensionless Reynolds stress anisotropy tensor $b_{ij} = \overline{u'_i u'_j} / \overline{u'_k u'_k} - \frac{1}{3}\delta_{ij}$. According to tensorial analysis, a symmetric second-order tensor can be diagonalized, and the eigenvalues are real numbers. The characteristic polynomial can be obtained
\begin{equation}
  \vert\mathbf{b} - \lambda \mathbf{I}\vert = \lambda^3 - \Psi_1^\mathbf{b}\lambda^2 + \Psi_2^\mathbf{b}\lambda - \Psi_3^\mathbf{b} = 0,
\end{equation}
where $\lambda$ is the eigenvalue of $b_{ij}$, in continuum mechanics, $\lambda$ also represents the principal stress. $\Psi_1^\mathbf{b}$, $\Psi_2^\mathbf{b}$, and $\Psi_3^\mathbf{b}$ are first, second, and third primary invariants of $b_{ij}$, respectively.
\begin{subequations}
  \begin{equation}
    \Psi_1^\mathbf{b} = tr\left(\mathbf{b}\right) = b_{ii} = 0, \label{54a}
  \end{equation}
  \begin{equation}
    \Psi_2^\mathbf{b} = \frac{1}{2}\left(b_{ii}b_{jj} - b_{ij}b_{ij}\right) = -\frac{1}{2}b_{ij}b_{ij}, \label{54b}
  \end{equation}
  \begin{equation}
    \Psi_3^\mathbf{b} = \vert \mathbf{b} \vert. \label{54c}
  \end{equation}
\end{subequations}
After the diagonalization of $b_{ij}$, we can get the dimensionless Reynolds stress anisotropy tensor in principle axes, which only contain the normal stress
\begin{equation}
    \widetilde{b_{ij}} = 
    \begin{pmatrix}
      \lambda_1 &  & \\
      & \lambda_2 & \\
      & & -\left(\lambda_1 + \lambda_2\right) 
    \end{pmatrix}. \label{55}
\end{equation}
Lumley \cite{Lumley1978} defined the parameter $\eta$ and $\xi$, according to the properties of $\widetilde{b_{ij}}$, we can derive the following relations
\begin{subequations}
  \begin{equation}
    6\eta ^ 2 = -2 I_2 = b_{ij}b_{ji} \Rightarrow \eta^2 = \frac{1}{3}\left(\lambda_1^2 + \lambda_1\lambda_2 + \lambda_2^2\right), \label{56a}
  \end{equation}
  \begin{equation}
    6 \xi^3 = 3I_3 = b_{ij}b_{jk}b_{ki} \Rightarrow \xi^3 = -\frac{1}{2}\lambda_1\lambda_2\left(\lambda_1+\lambda_2\right). \label{56b}
  \end{equation}
\end{subequations}
By considering a series of simplified turbulence conditions with Equation \eqref{56a} and \eqref{56b}, the Lumley triangle can be obtained. In other words, if the characteristic quantities of dimensionless Reynolds stress anisotropy tensor could not be wrapped by the Lumley triangle, it means that this turbulence model is not realizable. For $k-\varepsilon-\zeta$ model, characteristic quantities can be derived as
\begin{subequations}
  \begin{equation}
    6\eta ^ 2 = b_{ij}b_{ji} = \frac{\left(\zeta_{ijkl}\frac{\partial u_l}{\partial x_k}\right) \cdot \left(\zeta_{jikl}\frac{\partial u_l}{\partial x_k}\right)}{4k^2}, \label{57a}
  \end{equation}
  \begin{equation}
    6 \xi^3 = b_{ij}b_{jk}b_{ki} = - \frac{\left(\zeta_{ijmn}\frac{\partial u_n}{\partial x_m}\right) \cdot \left(\zeta_{jkmn}\frac{\partial u_n}{\partial x_m}\right) \cdot \left(\zeta_{kimn}\frac{\partial u_n}{\partial x_m}\right)}{8k^3}. \label{57b}
  \end{equation}
\end{subequations}
Thus, the results of the dimensionless Reynolds stress anisotropy tensor for $k-\varepsilon-\zeta$ model and standard $k-\varepsilon$ model is obtained, which are shown in Figure \ref{figure 1}. For characteristic quantities $\eta$ and $\xi$ of $k-\varepsilon-\zeta$ model and standard $k-\varepsilon$ model are both in the Lumley triangle. It means that $k-\varepsilon-\zeta$ model and standard $k-\varepsilon$ model are both realizable. However, for turbulent flow in the two-dimensional straight channel, the trajectory of the characteristic quantities of $k-\varepsilon-\zeta$ model is along the $\eta=\xi$, which is in good agreement with DNS channel flow \cite{Kim1987}. 

For a fully-developed straight channel turbulence, it is obvious that $\frac{\partial u_1}{\partial x_1} = 0$ and $\frac{\partial u_2}{\partial x_2} = 0$. Therefore, standard $k-\varepsilon$ model, which adopts the linear EVM constitutive, cannot capture the differences between $\overline{u'_1 u'_1}$ and $\overline{u'_2 u'_2}$ because linear EVM indicates that the turbulence is isotrpic, and the trajectory will pass through $\left(0, 0\right)$ in Lumley triangle. Hence, from Figure \ref{figure 1} we can know that $k-\varepsilon-\zeta$ model is superior to the standard $k-\varepsilon$ model in describing the characteristics of turbulence. 

\begin{figure}[H]
  \centering
  \subfloat[Components of $\zeta_{1111}, \zeta_{1211}$, and $\zeta_{2211}$]{%
  \resizebox*{7cm}{!}{\includegraphics{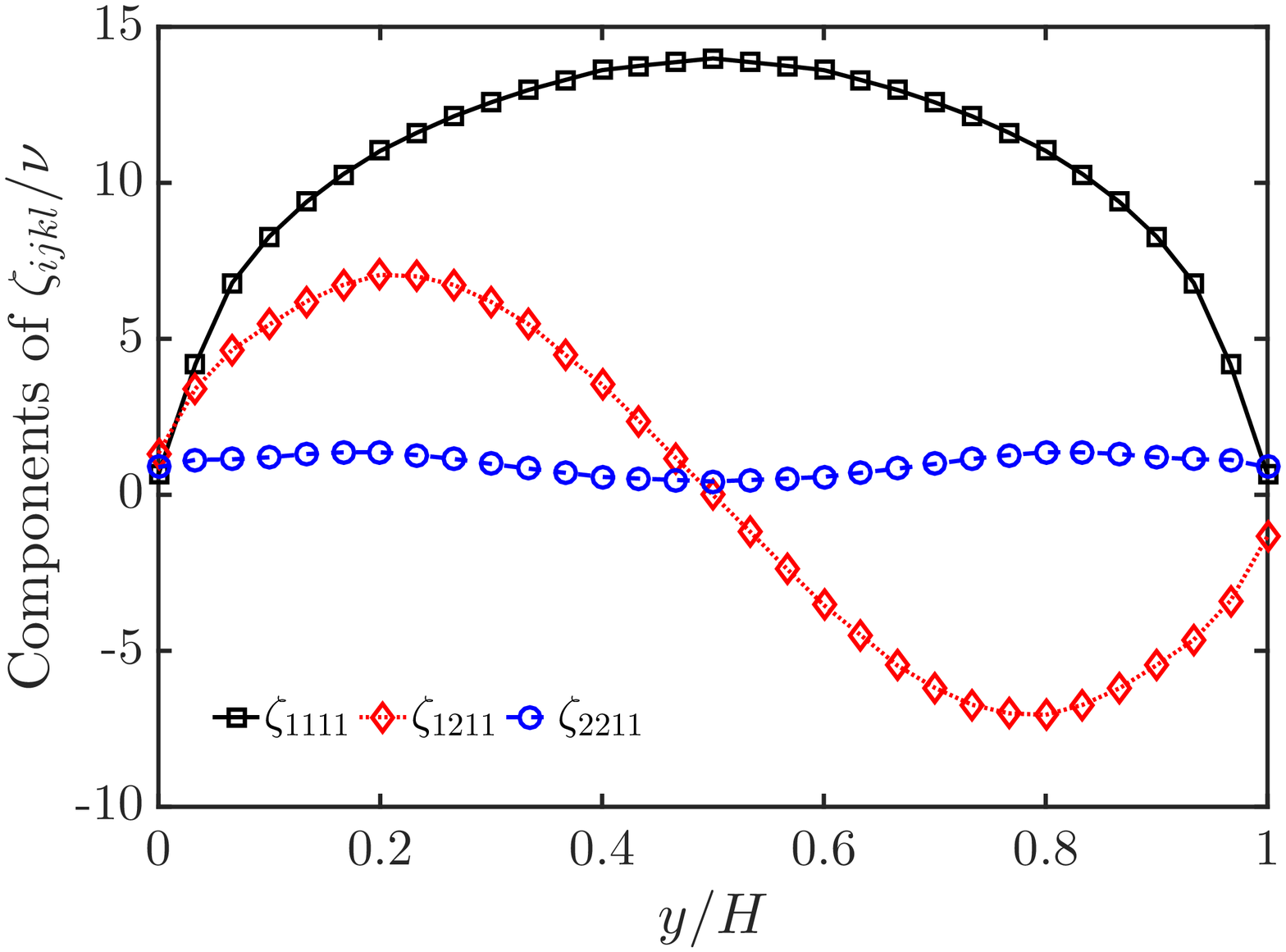}}}\hspace{5pt}
  \subfloat[Components of $\zeta_{1112}, \zeta_{1212}$, and $\zeta_{2212}$]{%
  \resizebox*{7cm}{!}{\includegraphics{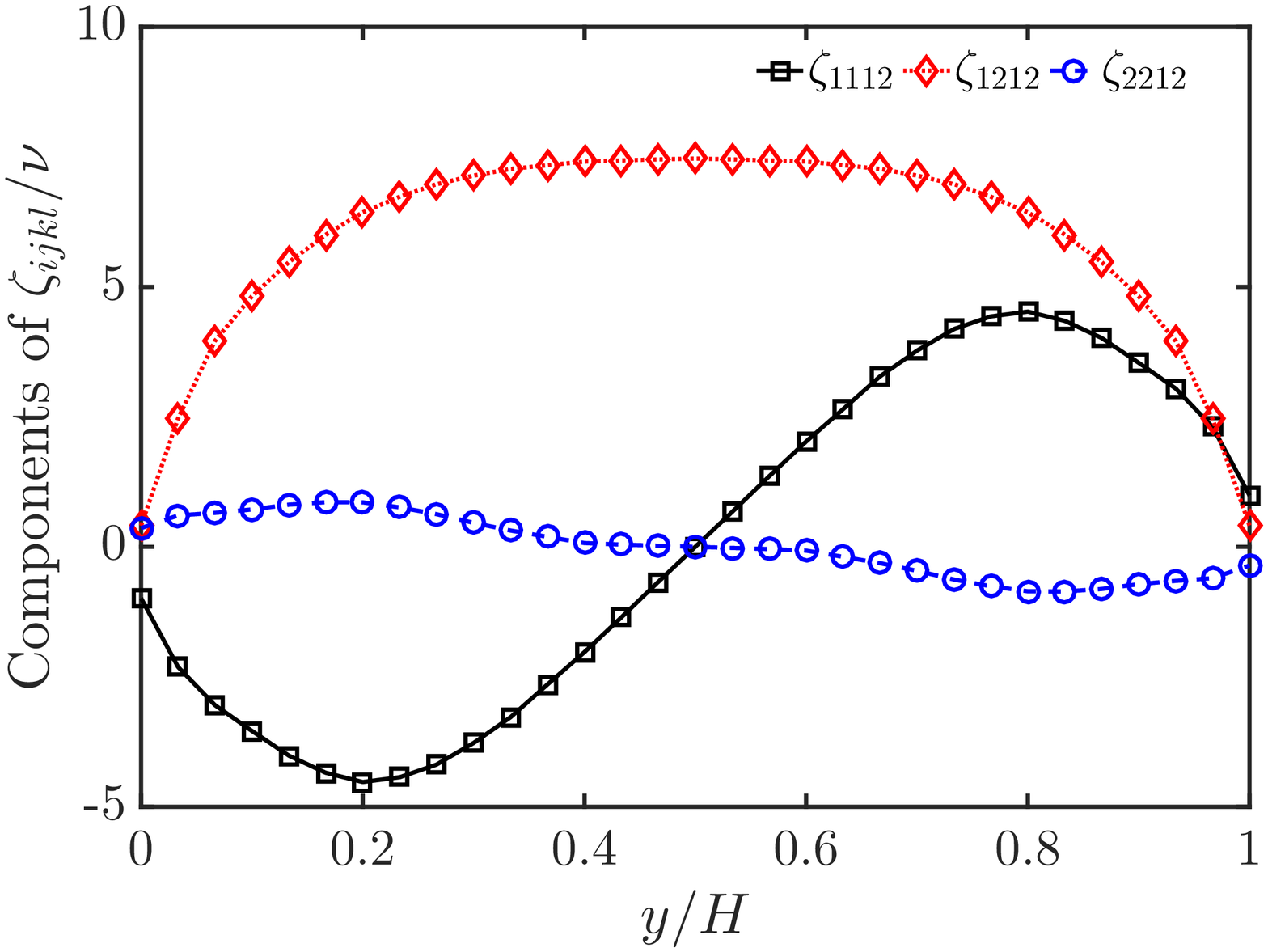}}}
  \qquad

  \subfloat[Components of $\zeta_{1121}, \zeta_{1221}$, and $\zeta_{2221}$]{%
  \resizebox*{7cm}{!}{\includegraphics{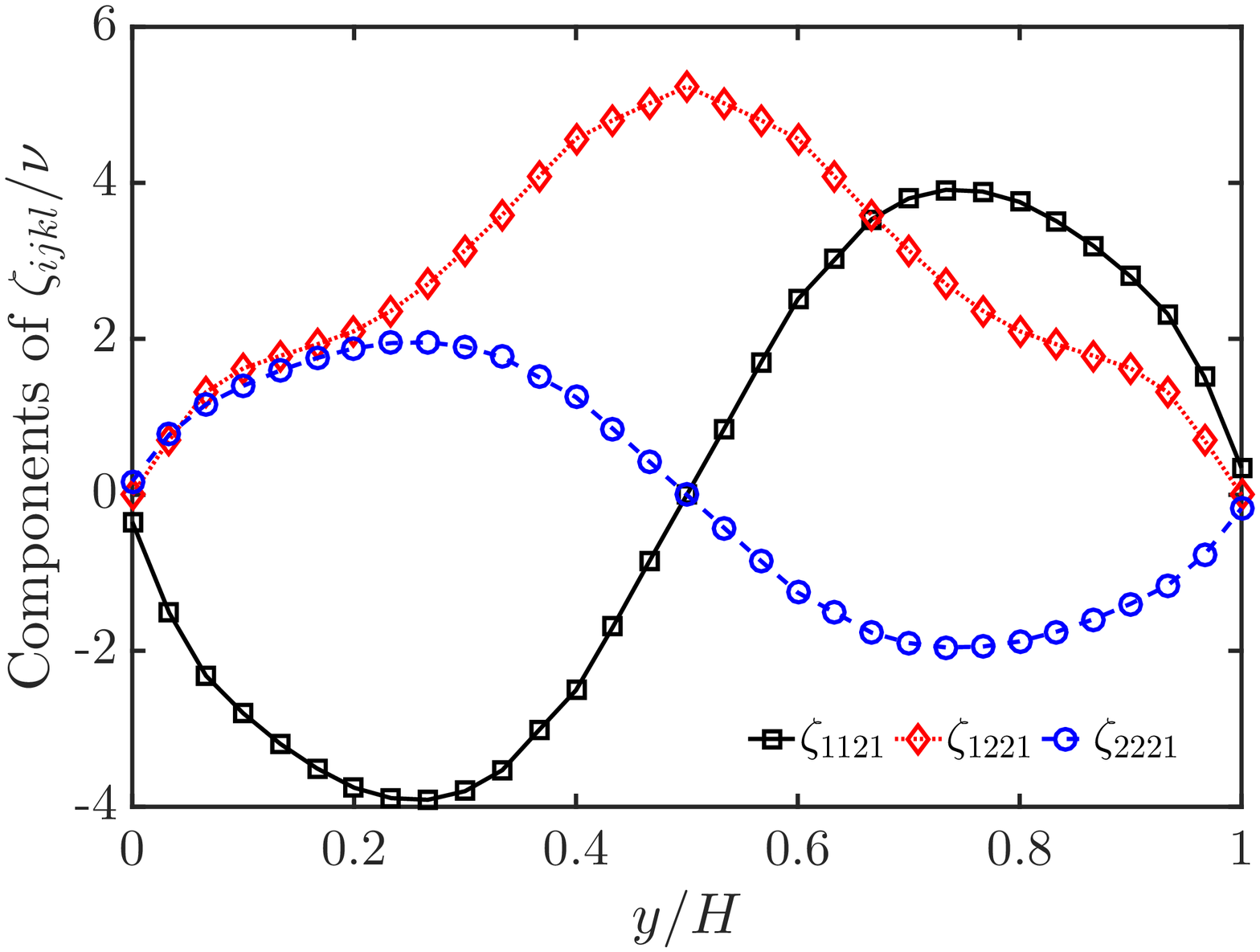}}}
  \subfloat[Components of $\zeta_{1122}, \zeta_{1222}$, and $\zeta_{2222}$]{%
  \resizebox*{7cm}{!}{\includegraphics{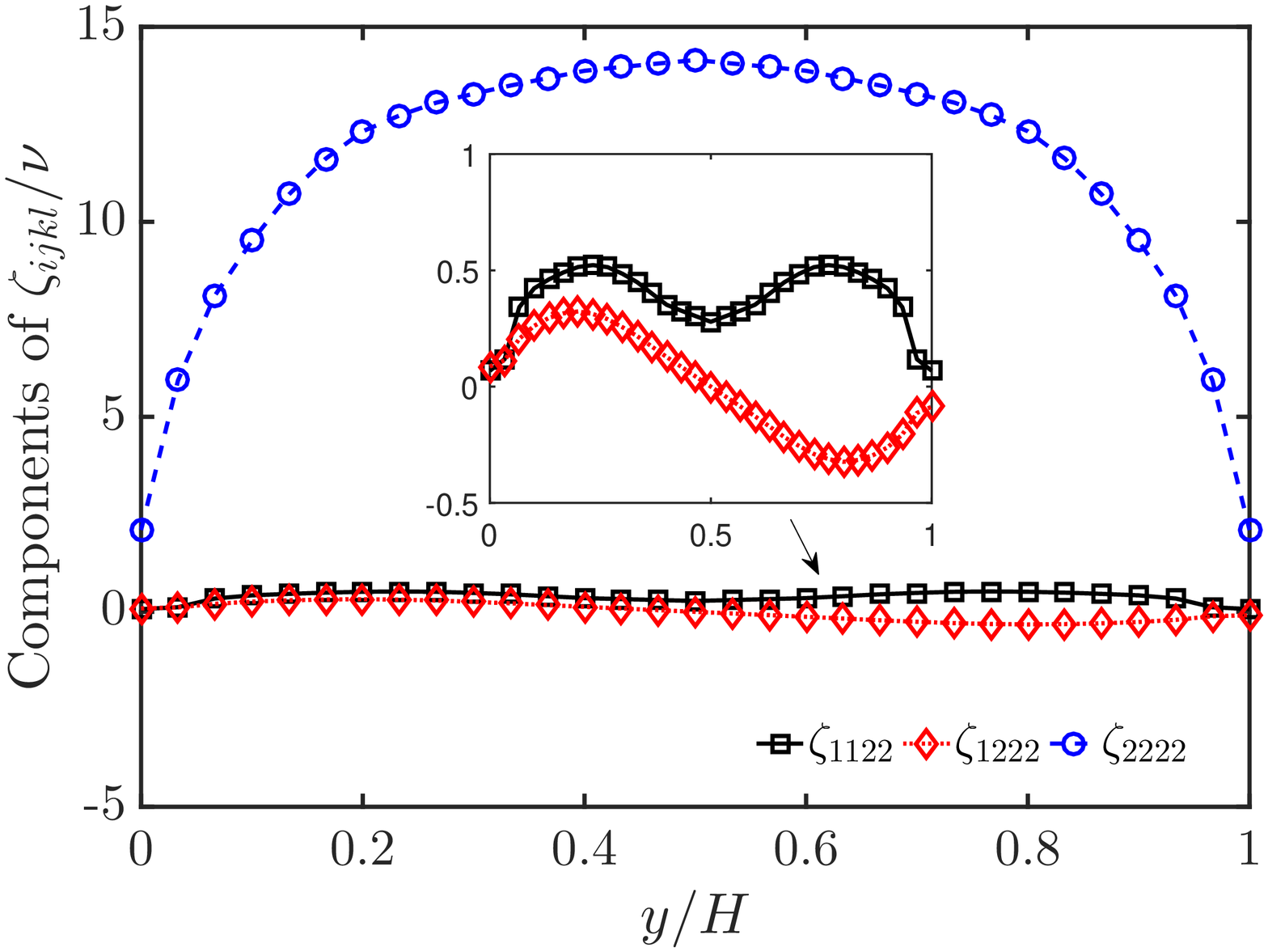}}}
  \caption{Components of high-order eddy viscosity tensor with $k-\varepsilon-\zeta$ model calculation. Horizontal ordinates and longitudinal ordinates are have been dimensionless.} \label{figure 2}
\end{figure}

By solving the transport equation of $\zeta_{ijkl}$, we also can get the distribution of each component of high-order eddy viscosity tensor. Figure \ref{figure 2} shows the numerical results of 12 independent components of high-order eddy viscosity tensor. From a numerical point of view, we can find that $\zeta_{1111}$ and $\zeta_{2222}$ play a major role in Reynolds viscous stress tensor, which $\zeta_{1111}$ acts on $R_{11}^V$, and $\zeta_{2222}$ acts on $R_{22}^V$.

We also can see that 12 independent components can be divided into two parts, the axisymmetric part, and the centrosymmetric part about the center of the straight channel. The main components of Reynolds viscous stress can be written in
\begin{subequations}
  \begin{equation}
    R_{11}^V = \underbrace{\zeta_{1111}\frac{\partial u_1}{\partial x_1} + \zeta_{1122}\frac{\partial u_2}{\partial x_2}}_{axisymmetric} + \underbrace{\zeta_{1112}\frac{\partial u_2}{\partial x_1} + \zeta_{1121}\frac{\partial u_1}{\partial x_2}}_{centrosymmetric}, \label{58a}
  \end{equation}
  \begin{equation}
    R_{12}^V = \underbrace{\zeta_{1212}\frac{\partial u_2}{\partial x_1} + \zeta_{1221}\frac{\partial u_1}{\partial x_2}}_{axisymmetric} + \underbrace{\zeta_{1211}\frac{\partial u_1}{\partial x_1} + \zeta_{1222}\frac{\partial u_2}{\partial x_2}}_{centrosymmetric}, \label{58b}
  \end{equation}
  \begin{equation}
    R_{22}^V = \underbrace{\zeta_{2211}\frac{\partial u_1}{\partial x_1} + \zeta_{2222}\frac{\partial u_2}{\partial x_2}}_{axisymmetric} + \underbrace{\zeta_{2212}\frac{\partial u_2}{\partial x_1} + \zeta_{2221}\frac{\partial u_2}{\partial x_1}}_{centrosymmetric}, \label{58c}
  \end{equation}
\end{subequations}
As mentioned above, for two-dimensional fully-developed turbulence in the straight channel, $\frac{\partial u_1}{\partial x_1} = 0$ and $\frac{\partial u_2}{\partial x_2} = 0$. Therefore, axisymmetric parts of $R_{11}^V$ and $R_{22}^V$ are nearly zero, and centrosymmetric parts reflect the anisotropy of Reynolds stress. For Reynolds viscous shear stress $R_{12}^V$, the main part is the axisymmetric part, and the centrosymmetric part is nearly zero. It can be seen from the value of each component that, for Reynolds viscous stress, the axisymmetric part is the main part, and the centrosymmetric part is more likely to play a regulatory role. 

For a complex two-dimensional turbulent flow, the flow in an asymmetric planar diffuser will be adopted in this article to make the numerical simulation of different RANS models. Turbulent flow in the asymmetric planar diffuser is a benchmark for Large Eddy Simulation (LES) because of the presence of an adverse pressure gradient, and the formation of an unsteady separation recirculation in the expansion section of the diffuser, which may carry out many challenges to LES prediction \cite{Kaltenbach1999}. Many simulations based on LES are carried out for the asymmetric planar diffuser \cite{Hajaali2022,Kaltenbach1999,Tang2019}. The turbulent flow in the diffuser contains the behaviors of adverse pressure gradient of flow and anisotropy and relaxation of Reynolds stress, which are also critical to the validation of RANS models. Obi et al. \cite{Obi1993} made experimental and numerical researches on the diffuser in a two-dimensional turbulent flow. The LDV measurement of the turbulent separating flow in an asymmetric diffuser was used to capture the behavior of the turbulent flow. Obi et al. \cite{Obi1993} demonstrated that standard $k-\varepsilon$ model fails to predict the separation of the flow. In this article, $k-\varepsilon-\zeta$ model will make the simulation on two-dimensional turbulent flow in the diffuser, investigation of the transportation of high-order eddy viscosity tensor will also be carried out.

\begin{figure}[H]
  \centering
  \includegraphics[width=1.0\linewidth]{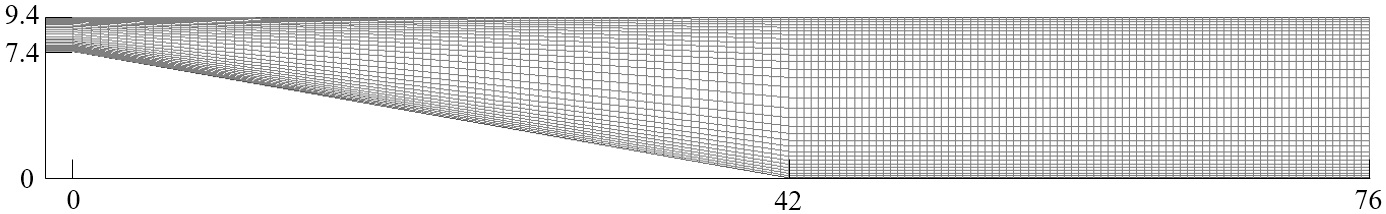}
  \caption{Computational domain and mesh of the two-dimensional diffuser. Only the expansion part of the diffuser is plotted. The geometric dimensions are from Obi et al. \cite{Obi1993} experiments, and the units are in cm.}
  \label{figure 3}
\end{figure}

Figure \ref{figure 3} shows the computational domain and mesh of the expansion part of the diffuser, and the geometric dimensions are referred to the Obi et al. \cite{Obi1993} experiments. The height of the inlet channel is $H=2cm$, and the length of the inlet channel is $100H$ to ensure the fully-developed two-dimensional turbulent channel flow at $Re=2.0\times 10^4$. Obi et al. \cite{Obi1993} also examined the two-dimensionality of the flow field, and the difference in mean velocity profile in the spanwise direction was verified to be less than 5\% over 90\% and 60\% of the channel span at the inlet and outlet plane. In this article, the inlet velocity $U_{in}$ is $2m/s$, and the kinematic viscosity $\nu$ is $10^{-6}$ to ensure the channel flow at $Re=2.0\times 10^4$.

\begin{figure}[H]
  \centering
  \subfloat[Streamwise mean velocity calculated by $k-\varepsilon-\zeta$ model]{%
  \resizebox*{12cm}{!}{\includegraphics{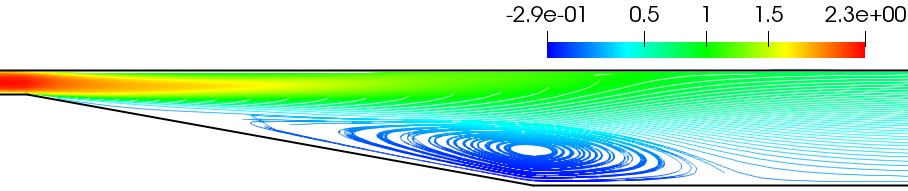}}}\vspace{5pt}
  \qquad
  \subfloat[Streamwise mean velocity calculated by standard $k-\varepsilon$ model]{%
  \resizebox*{12cm}{!}{\includegraphics{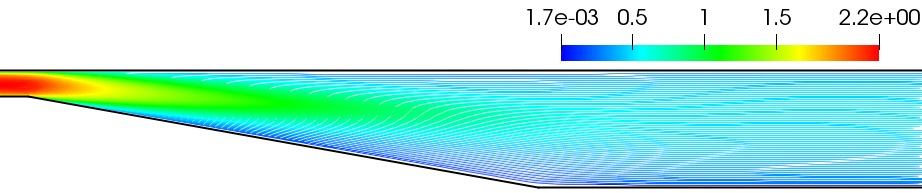}}}
  \caption{Streamline and streamwise mean velocity calculated by $k-\varepsilon-\zeta$ model and standard $k-\varepsilon$ model. (a) The result of $k-\varepsilon-\zeta$ model, which can predict the flow separation in the expansion section of the diffuser because of the capture of the anisotropy and relaxation of Reynolds stress. (b). The result of standard $k-\varepsilon$, which fails to predict the flow separation in the expansion section of the diffuser. Both units are $m/s$ in (a) and (b).} \label{figure 4}
\end{figure}

Figure \ref{figure 4} shows the simulation results which are calculated by OpenFOAM for $k-\varepsilon-\zeta$ model and standard $k-\varepsilon$ model. From the results, it is obvious that $k-\varepsilon-\zeta$ model can predict the separation bubble, and standard $k-\varepsilon$ model fails to predict the separation bubble. These phenomena are in the agreement with Obi’s numerical results. Therefore, $k-\varepsilon-\zeta$ model can capture the anisotropy and relaxation of Reynolds stress, while standard $k-\varepsilon$ model cannot. The experimentally obtained flow detachment and reattachment points are $11H$ and $26H$, respectively \cite{Obi1993}. For $k-\varepsilon-\zeta$ model, the calculation results obtain the flow detachment and reattachment points are nearly $6.16H$ and $27H$, respectively. The prediction of the separation bubble is in good agreement with the experimental measurements. Kaltenbach et al. \cite{Kaltenbach1999} carried out the three-dimensional LES simulation for the planar asymmetric diffuser, the detachment and reattachment points are also investigated. Crawford and Birk \cite{Crawford2015} calculated the asymmetric planar diffuser based on the $v^2-f$ turbulence model \cite{Durbin1995} of ANSYS Fluent to study the flow characteristics in two-dimensional turbulent flow, which also contained the detachment and reattachment points in the expansion section of the diffuser. All the results are summarized in Table 1.

\begin{table}[H]
  \tbl{Comparison of Detachment and Reattachment points and bubble length between different turbulence models and experimental data.}
  {\begin{tabular}{lccc} \toprule
   Turbulence model or experimental data & Detachment & Reattachment & Bubble length  \\ \midrule
   Obi Experiment \cite{Obi1993} & $11H$ & $26H$ & $15H$  \\
   $k-\varepsilon-\zeta$ model & $6.16$ & $27$ & $20.84$  \\
   LES \cite{Kaltenbach1999} & $6H$ & $27.5H$ & $21.5H$  \\
   $v^2-f$ \cite{Crawford2015} & $6.15H$ & $27.6H$ & $21.45H$  \\ \bottomrule
  \end{tabular}}
  \tabnote{Note: All parameters are in the width of the inlet channel $H=0.02m$}
\label{sample-table}
\end{table}

\begin{figure}[H]
  \centering
  \subfloat[Dimensionless streamwise mean velocity at $13.2H$]{%
  \resizebox*{7cm}{!}{\includegraphics{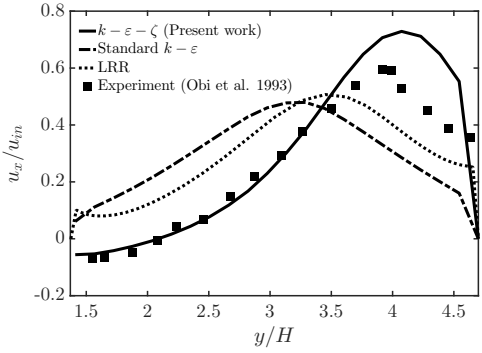}}}\hspace{5pt}
  \subfloat[Dimensionless Reynolds shear stress at $13.2H$]{%
  \resizebox*{7cm}{!}{\includegraphics{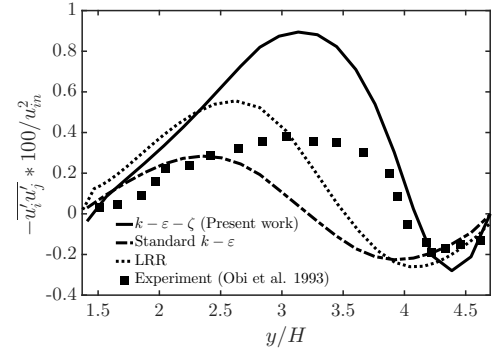}}}
  \qquad

  \subfloat[Dimensionless streamwise mean velocity at $19.2H$]{%
  \resizebox*{7cm}{!}{\includegraphics{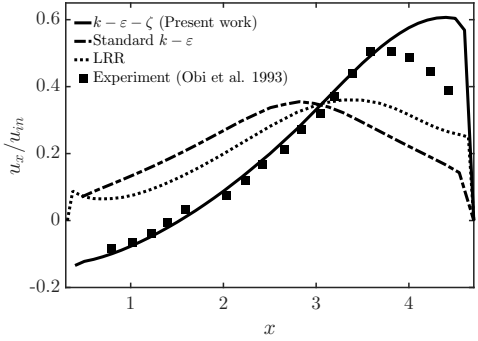}}}
  \subfloat[Dimensionless Reynolds shear stress at $19.2H$]{%
  \resizebox*{7cm}{!}{\includegraphics{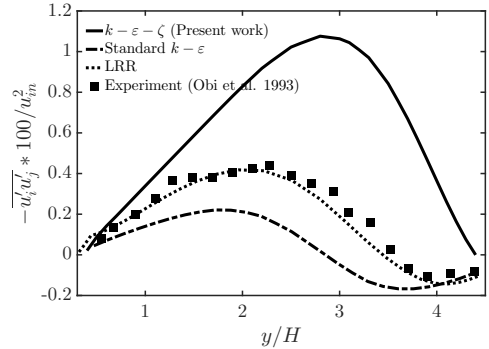}}}
  \caption{Results of dimensionless streamwise mean velocity and dimensionless Reynolds shear stress at $13.2H$ and $19.2H$. Solid line represents the results of $k-\varepsilon-\zeta$ model. Dotted line represents the results of standard $k-\varepsilon$ model. Dashed line represents the results of LRR model. Dot represents the results of experiment.} \label{figure 5}
\end{figure}

To better analyze the turbulent flow characteristics, we carry out quantitative analysis, which compares with the results of Obi et al. \cite{Obi1993} experimental measurements. Figure \ref{figure 5} shows the results of dimensionless streamwise mean velocity and dimensionless Reynolds shear stress at $13.2H$ and $19.2H$ for $k-\varepsilon-\zeta$ model, standard $k-\varepsilon$ model, and LRR model, respectively. From Figure \ref{figure 3}, we can know that $13.2H$ (26.4 cm) and $19.2H$ (38.4 cm) are both at the expansion section of the asymmetric planar diffuser, which is a strict region for turbulent flow. Obi et al. \cite{Obi1993} also carried out a numerical simulation using LRR model, hence we also add a simulation with LRR model to calculate the two-dimensional planar asymmetric diffuser in this article, to make the comparison and analysis with $k-\varepsilon-\zeta$ model and standard $k-\varepsilon$ model. 

From the calculation results, neither LRR model nor standard $k-\varepsilon$ model can simulate the separation bubble in the expansion section. $k-\varepsilon-\zeta$ model can simulate the separation bubble, and the results of streamwise mean velocity at $13.2H$ and $19.2H$ are in good agreement with the experimental measurements. At $13.2H$ in the expansion section, the prediction accuracy of the streamwise mean velocity of $k-\varepsilon-\zeta$ model is increased by nearly 300\% when compares to standard $k-\varepsilon$ model, and nearly 200\% when compares to LRR model. And at $19.2H$ in the expansion section, the prediction accuracy of the streamwise mean velocity of $k-\varepsilon-\zeta$ model is increased by nearly 200\% and 100\% when compared with standard $k-\varepsilon$ model and LRR model, respectively. However, the streamwise mean velocity of   model is higher than the experimental measurement at the upper wall of the asymmetric planar diffuser, because of the numerical approximation of the production term of turbulent kinetic energy in Equation \eqref{52b}. For the lower wall, the velocity gradient near the wall is not large due to the low velocity in the separation bubble region, hence $k-\varepsilon-\zeta$ model can simulate the streamwise mean velocity well. When near the upper wall region, the velocity is much larger than the separation bubble region, hence the velocity gradient is much larger than the separation bubble region, the effect of the numerical approximation in Equation \eqref{52b} is obvious and the numerical deviation appears. 

For Reynolds shear stress at $13.2H$ and $19.2H$ in the expansion section, we can find that standard $k-\varepsilon$ model always underestimates the Reynolds shear stress, and $k-\varepsilon-\zeta$ model always overestimates the Reynolds shear stress, which are shown in Figure \ref{figure 5}.(b) and (d). Only the LRR model accurately estimates the Reynolds shear stress at $19.2H$. The main reason for the overestimation of $k-\varepsilon-\zeta$ model is the numerical approximation in Equation \eqref{52b}, which will overestimate the turbulent kinetic energy and Reynolds shear stress. However, from the overall calculation results, we can see that the $k-\varepsilon-\zeta$ model can preliminarily reveal the turbulent flow characteristics of the two-dimensional asymmetric planar diffuser. 

\begin{figure}[H]
  \centering
  \subfloat[$\zeta_{1111}$]{%
  \resizebox*{7cm}{!}{\includegraphics{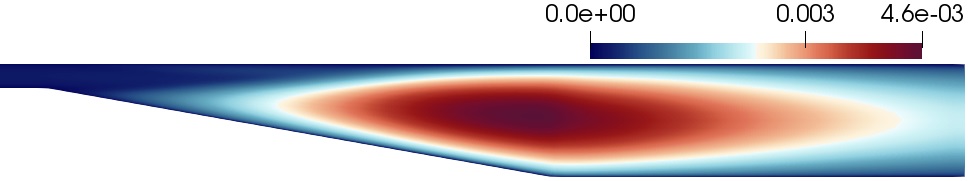}}}\hspace{5pt}
  \subfloat[$\zeta_{1112}$]{%
  \resizebox*{7cm}{!}{\includegraphics{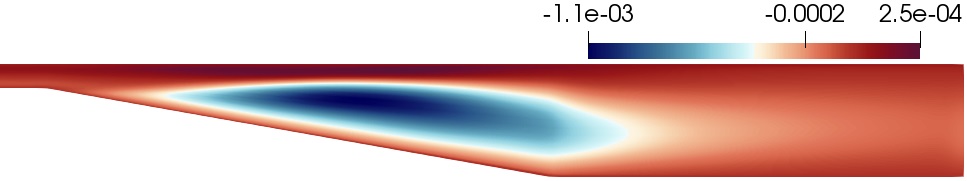}}}
  \qquad

  \subfloat[$\zeta_{1121}$]{%
  \resizebox*{7cm}{!}{\includegraphics{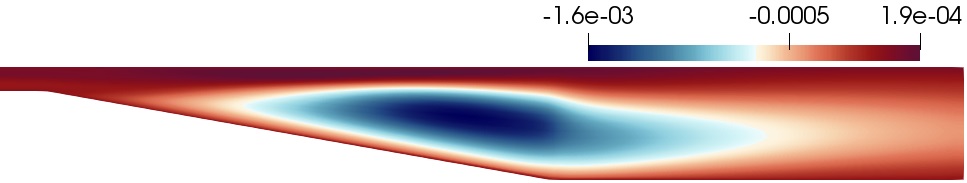}}}\hspace{5pt}
  \subfloat[$\zeta_{1122}$]{%
  \resizebox*{7cm}{!}{\includegraphics{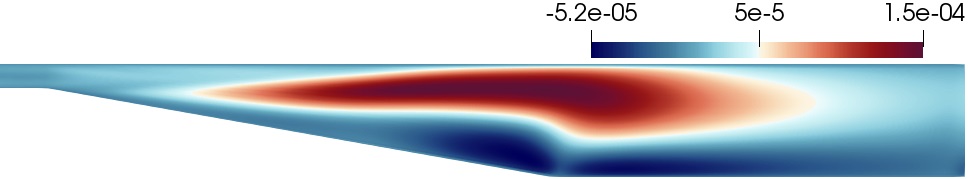}}}
  \qquad

  \subfloat[$\zeta_{1211}$]{%
  \resizebox*{7cm}{!}{\includegraphics{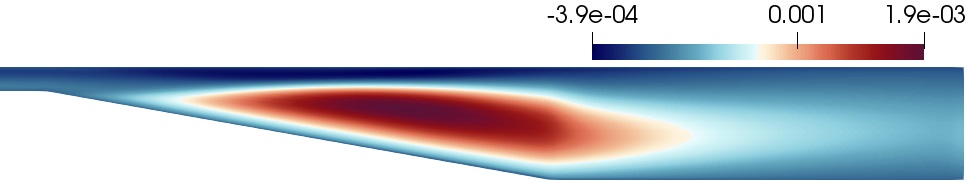}}}\hspace{5pt}
  \subfloat[$\zeta_{1212}$]{%
  \resizebox*{7cm}{!}{\includegraphics{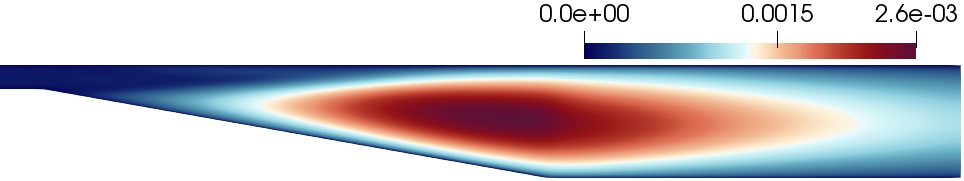}}}
  \qquad

  \subfloat[$\zeta_{1221}$]{%
  \resizebox*{7cm}{!}{\includegraphics{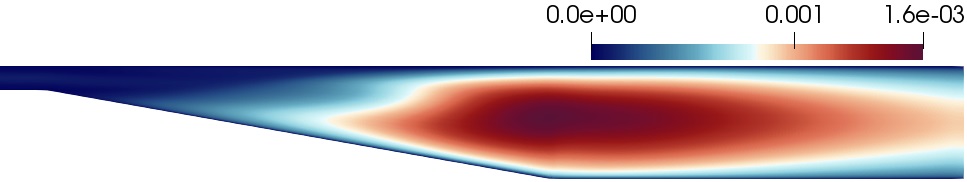}}}\hspace{5pt}
  \subfloat[$\zeta_{1222}$]{%
  \resizebox*{7cm}{!}{\includegraphics{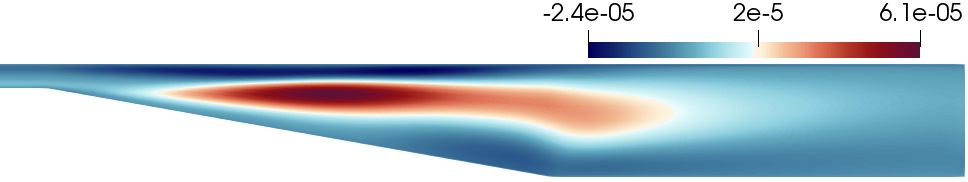}}}
  \qquad

  \subfloat[$\zeta_{2211}$]{%
  \resizebox*{7cm}{!}{\includegraphics{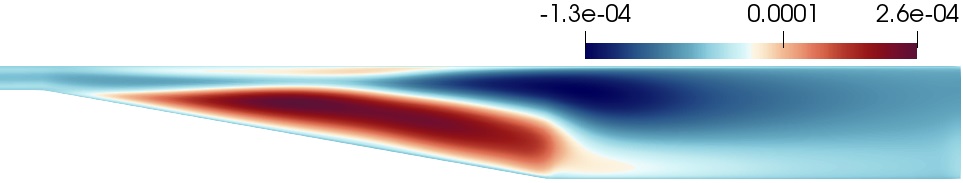}}}\hspace{5pt}
  \subfloat[$\zeta_{2212}$]{%
  \resizebox*{7cm}{!}{\includegraphics{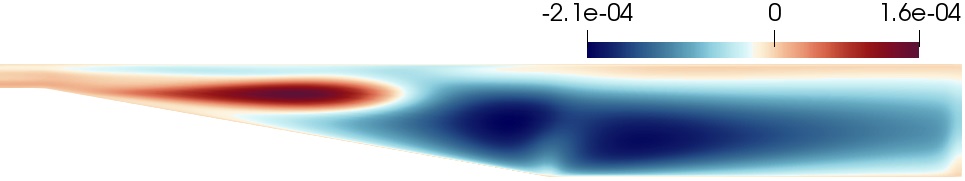}}}
  \qquad

  \subfloat[$\zeta_{2221}$]{%
  \resizebox*{7cm}{!}{\includegraphics{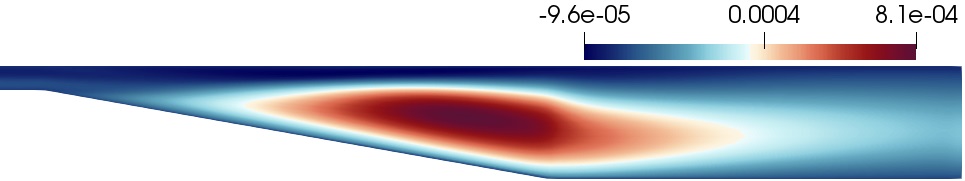}}}\hspace{5pt}
  \subfloat[$\zeta_{2222}$]{%
  \resizebox*{7cm}{!}{\includegraphics{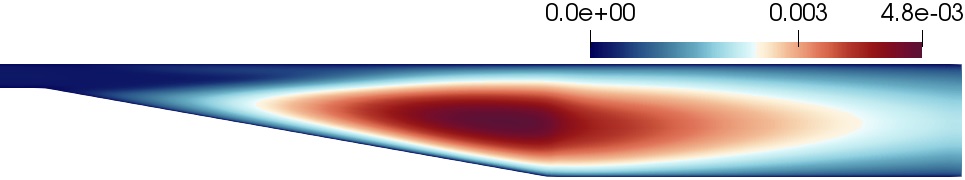}}}

  \caption{Distribution of independent components of high-order eddy viscosity tensor which are calculated by $k-\varepsilon-\zeta$ model. All the units are in $m^2/s^2$} \label{figure 6}
\end{figure}

By solving the conserving Equation \eqref{48}, high-order eddy viscosity tensor $\zeta_{ijkl}$ can be calculated, which contains 12 independent components. Figure \ref{figure 6} shows the results of 12 independent components of high-order eddy viscosity tensor of complex turbulent flow in an asymmetric planar diffuser. The variation of the distribution of $\zeta_{ijkl}$ mainly starts near the flow separation point. And the positive and negative signs of the components of $\zeta_{ijkl}$ will change after the flow separation in the expansion section. 

For an asymmetric planar diffuser, numerical analysis \cite{Crawford2015,Durbin1995} shows that the low-Re $v^2-f$ model, which considers the curvature correction and the near-wall flow characteristics, has a good performance and quite a high degree of accuracy. For low-Re models, such as $v^2-f$ model, $k-\omega$ SST model \cite{Menter1993,Menter1994}, they capture more accurately the changes of geometric curvature and the changes of velocity gradient near the wall. Therefore, a large part of the reason why they capture the separated flow is that they capture that the $\tau_{\shortparallel} = \mu \frac{\partial u_{\shortparallel}}{\partial x_{\shortparallel}} = 0$ near the wall due to the adverse pressure gradient is 0, in which $\tau_{\shortparallel}$ is the shear stress parallel to the wall, $\frac{\partial u_{\shortparallel}}{\partial x_{\shortparallel}}$ is the normal gradient of the velocity which is parallel to the wall. 

However, for the high-Re models that need to use wall functions to approximate the near-wall flow characteristics, the capture of geometric curvature and the near-wall flow behavior is weaker than that of low-Re models. Due to $k-\varepsilon-\zeta$ model is derived from LRR model, hence $k-\varepsilon-\zeta$ model is also a high-Re model. It means that the capture of the separation bubble in the expansion region of the diffuser is based on the properties of Reynolds stress which are characterized by the turbulence model, rather than the analysis of flow characteristics near the wall like the low-Re model. The same phenomenon also can be observed when using nonlinear EVM to calculate the internal flow in a square duct, that the induced vortices due to the anisotropy property of Reynolds stress occur on the cross-section \cite{Speziale1987}. 

\section{Discussion and conclusion}
\subsection{2D-3C turbulence structure and EV-RSM}

In this article, the development of $k-\varepsilon-\zeta$ model is carried out under the 2D-3C turbulence structure. To explicitly obtain the transport model of high-order eddy viscosity tensor, however, analogy from the linear EVM model, an assumption for constitutive relation is made in Equation \eqref{15}. It means that the Reynolds viscous normal stress on the z-axis is $R_{33}^V=0$, and Reynolds normal stress on the z-axis is $R_{33} = \frac{2}{3}k$. However, for the second-order moment model, when a two-dimensional turbulence flow in a 2D-3C turbulence structure is established, the Reynolds normal stress on the z-axis $R_{33}$ still needs to be solved by a transport equation. This will result in $R_{33}$ not equal to $\frac{2}{3}k$. The transport equation of $R_{33}$ in 2D-3C turbulence structure is 
\begin{equation}
  \begin{aligned}
    \frac{\partial \overline{u'_3 u'_3}}{\partial t} + u_1\frac{\partial \overline{u'_3 u'_3}}{\partial x_1} + u_2\frac{\partial \overline{u'_3 u'_3}}{\partial x_2} = &-\frac{\partial \overline{u'_3 u'_3 u'_1}}{\partial x_1} - \frac{\partial \overline{u'_3 u'_3 u'_2}}{\partial x_2} \\
    &- \frac{\partial }{\partial x_1}\left(\nu\frac{\partial \overline{u'_3 u'_3}}{\partial x_1}\right) - \frac{\partial }{\partial x_2}\left(\nu\frac{\partial \overline{u'_3 u'_3}}{\partial x_2}\right) \\
    &-2\nu\overline{\frac{\partial u'_3}{\partial x_1} \cdot \frac{\partial u'_3}{\partial x_1} + \frac{\partial u'_3}{\partial x_2}\cdot \frac{\partial u'_3}{\partial x_2}}. \label{59}
  \end{aligned}
\end{equation}
This may contain the incompatibility between RSM (Reynolds Stress Model) and EV-RSM (Eddy Viscosity – Reynolds Stress Model). EV-RSM is a RANS turbulence model, which uses the Boussinesq hypothesis to maintain the simplicity and considers incorporating the second-order moment transport equation to improve the accuracy in describing the properties of Reynolds stress. It is obvious that $k-\varepsilon-\zeta$ model is an EV-RSM model, because this model adopts the Boussinesq hypothesis and is derived from the second-order moment model. Spalart-Allmaras one-equation model is also an EV-RSM model, the transport equation of eddy viscosity $\nu_t$ has a strict form, which can be built with the second-order moment model. In Spalart-Allmaras one-equation model, the constitutive relation is the linear EVM hypothesis, which is shown in Equation \eqref{7}. For incompressible turbulent flow, square both sides of Equation \eqref{7}, and simplify with the continuity equation, we can get
\begin{equation}
  \overline{u'_i u'_j} \cdot \overline{u'_i u'_j} = \frac{4}{3}k^2 + 4\nu_t^2 S_{kl}S_{kl}. \label{60}
\end{equation}
Make the material differentiation of Equation \eqref{60}, we can get
\begin{equation}
  \frac{D}{Dt}\left(\overline{u'_i u'_j} \cdot \overline{u'_i u'_j}\right) = \frac{8}{3}k\frac{Dk}{Dt} + 4\nu_t^2\frac{D}{Dt}\left(S_{kl}S_{kl}\right) + 8S_{kl}S_{kl}\nu_t\frac{D\nu_t}{Dt}. \label{61}
\end{equation}
Substitute the constitutive relation of linear EVM in Equation \eqref{7}, and make the simplification with the continuity equation, we can get
\begin{equation}
  -\frac{D}{Dt}\left(2\nu_t S_{ij}\overline{u'_i u'_j}\right) = 4\nu_t^2\frac{D}{Dt}\left(S_{kl}S_{kl}\right) + 8S_{kl}S_{kl}\nu_t\frac{D\nu_t}{Dt}. \label{62}
\end{equation}
With the mathematical transformation, the left-hand side term can be transformed into
\begin{equation}
  -\frac{D}{Dt}\left(2\nu_t S_{ij}\overline{u'_i u'_j}\right) = -\overline{u'_i u'_j}\frac{D}{Dt}\left(2\nu_t S_{ij}\right) - 2\nu_t S_{ij}\frac{D \overline{u'_i u'_j}}{Dt}. \label{63}
\end{equation}
Substitute the constitutive relation of linear EVM in Equation \eqref{7}, we can get
\begin{equation}
  \begin{aligned}
    -\frac{D}{Dt}\left(2\nu_t S_{ij}\overline{u'_i u'_j}\right) &= 2\nu_t S_{ij}\frac{D}{Dt}\left(2 \nu_t S_{ij}\right) - 2\nu_t S_{ij}\frac{D \overline{u'_i u'_j}}{Dt} \\
    &=-2\nu_t S_{ij}\frac{D \overline{u'_i u'_j}}{Dt} - 2\nu_t S_{ij}\frac{D \overline{u'_i u'_j}}{Dt} \\
    &=-4\nu_t S_{ij}\frac{D \overline{u'_i u'_j}}{Dt}. \label{64}
  \end{aligned}
\end{equation}
Hence Equation \eqref{62} can be converted to 
\begin{equation}
  -4\nu_t S_{ij}\frac{D \overline{u'_i u'_j}}{Dt} = 4\nu_t^2\frac{D}{Dt}\left(S_{kl}S_{kl}\right) + 8S_{kl}S_{kl}\nu_t\frac{D\nu_t}{Dt}. \label{65}
\end{equation}
We can find that $8S_{kl}S_{kl}\nu_t$ is a scalar, which is the reason why we square the linear EVM constitutive relation at first. Divide both sides of the Equation \eqref{65} by $8S_{kl}S_{kl}\nu_t$
\begin{equation}
  \frac{D \nu_t}{Dt} = -\frac{S_{ij}}{2S_{kl}S_{kl}}\frac{D \overline{u'_i u'_j}}{Dt} - \frac{\nu_t}{2S_{kl}S_{kl}}\frac{D}{Dt}\left(S_{ij}S_{ij}\right). \label{66}
\end{equation}
Hence, we can get the transport model of eddy viscosity $\nu_t$ in the strict form
\begin{equation}
  \begin{aligned}
    \frac{\partial \nu_t}{\partial t} + u_m \frac{\partial \nu_t}{\partial x_m} = &-\frac{S_{ij}}{2S_{kl}S_{kl}}\underbrace{\left(\frac{\partial \overline{u'_i u'_j}}{\partial t} + u_m \frac{\partial \overline{u'_i u'_j}}{\partial x_m}\right)}_{Second-order \ moment \ model} \\
    &- \frac{\nu_t}{2S_{kl}S_{kl}}\left[\frac{\partial \left(S_{ij}S_{ij}\right)}{\partial t} + u_m \frac{\partial \left(S_{ij}S_{ij}\right)}{\partial x_m}\right]. \label{67}
  \end{aligned}
\end{equation}
It can be seen that the strict form of transport model of eddy viscosity $\nu_t$ contains the second-order moment transport model. Therefore, theoretically, Spalart-Allmaras one-equation model also has incompatibility when calculating the two-dimensional turbulent flow in 2D-3C turbulence structure. Because from the constitutive relation of Spalart-Allmaras one-equation, the Reynolds viscous normal stress on the z-axis is $R_{33}^V=0$ and the Reynolds normal stress on the z-axis is $R_{33}=\frac{2}{3}k$, but the strict form transport model of eddy viscosity $\nu_t$ contains the second-order moment model, which needs to solve the transport Equation \eqref{59} of $\overline{u'_3 u'_3}$ in two-dimensional turbulent flow.

It is also not a good choice to directly model and solve Equation \eqref{67}. Because from Equation \eqref{67} we can know that if we want to get eddy viscosity $\nu_t$, we need to solve the second-order moment equation first. If we have already solved the second-order moment equation, the Reynolds stress is obtained, we do not need to solve other equations or make another hypothesis further. In 1994, Spalart and Allmaras \cite{Spalart1994} empirically developed the one-equation model, which is motivated by algebraic models such as the Baldwin-Lomax model \cite{Bladwin1978}, and the Johnson-King model \cite{Johnson1985}. Spalart and Allmaras considered the one-equation model of four nested operating conditions, from the simplest condition which applicable only to free shear flows to the most complete condition which applicable to viscous flows past solid bodies and with laminar regions. For each new operating condition, some semi-theoretical and semi-empirical source terms would be added to the transport equation of eddy viscosity $\nu_t$. By comparing with the experimental measurements, source terms and empirical constants in the transport model of eddy viscosity would be calibrated and updated. Therefore, Spalart-Allmaras one-equation model is obtained, which is successful in the field of aerodynamics with the small amount of calculation, stable convergence, and explanation of historical effect and relaxation of Reynolds stress. Based on the above analysis, we can find that the incompatibility between EV-RSM and the second-order moment model of two-dimensional turbulent flow under the 2D-3C turbulence structure can be reduced by introducing some empirical constants or source terms artificially depending on experimental measurements or DNS data calibration.

Motivated by Spalart-Allmaras one-equation model, an empirical damping scalar function $f\left(Re_t, \mathbf{u}, \nabla \mathbf{u}, \cdots\right)$ could be added on $P_k^{apx}$ in Equation \eqref{52b}, as 
\begin{equation}
  P_k^{apx} = f\left(Re_t, \mathbf{u}, \nabla \mathbf{u}, \cdots\right)\cdot \nu_t \left(\frac{\partial u_i}{\partial x_j} + \frac{\partial u_j}{\partial x_i}\right)\frac{\partial u_i}{\partial x_j}, \label{68}
\end{equation}
in which $Re_t = k^2 / \left(\nu \varepsilon\right)$ is the turbulent Reynolds number. We think that the damping function is the function of $Re_t$, $\mathbf{u}$, $\nabla \mathbf{u}$ and other physical quantities. Thus, we can fit the empirical damping function by experimental measurements or DNS data. The fitting function of $f\left(Re_t, \mathbf{u}, \nabla \mathbf{u}, \cdots\right)$ can be derived with machine learning or neural network in the future, which is not the focus of this article. Or to find a more stable numerical algorithm and iterative strategy to solve the large nonlinear equations of $k-\varepsilon-\zeta$ model without any numerical approximation and simplification. From an engineering perspective, the $k-\varepsilon-\zeta$ model developed in this article may not be practical, because this model contains a large number of nonlinear equations, and the computational complexity, convergence, and stability are urgent problems to be solved. However, $k-\varepsilon-\zeta$ model supplies a new method to study the high-order eddy viscosity tensor in two-dimensional turbulent flow with 2D-3C turbulence structure, which reveals the evolution law and some physical properties of high-order eddy viscosity tensor in the aspect of physics.

\subsection{Conclusion and outlook}
A framework of the transport model for high-order eddy viscosity tensor in 2D-3C turbulence structure has been developed which is named as $k-\varepsilon-\zeta$ model in this article. Starting from the constitutive relation of the Boussinesq hypothesis in RANS, by analogy with the constitutive of Navier-Stokes hypothesis, we obtain the constitutive relation in the form of high-order eddy viscosity tensor without simplification and approximation. Through the modeled second-order moment transport equation and turbulent kinetic energy transport equation, we can derive the transport equation of Reynolds viscous stress. Regarding the constitutive relation of high-order eddy viscosity tensor as a tensor function, we obtain the explicit transport equation of high-order eddy viscosity tensor through the derivative of the tensor function and mathematical transformation. The physical meaning of the source terms in the transport equation of high-order eddy viscosity tensor is also analyzed. For the transport equation of high-order eddy viscosity tensor, it contains the transient term, convection term, and source terms. We can figure out that high-order eddy viscosity tensor can be generated by the normal elastic deformation in the strain production term, and also can be generated by the change of time-averaging vorticity in the vorticity production term. The diffusion term also has been concluded in the transport equation to make the high-order eddy viscosity tensor isotropic. The dissipation term of high-order eddy viscosity tensor is obtained to characterize the dissipation behavior of the turbulent flow. In addition, a positive-definite source term also appears in the transport equation, when the components of high-order eddy viscosity tensor are positive, this source term acts as a source that will increase the components. When the components are negative, the positive-definite source acts as a sink which will decrease the components to prevent the CGT phenomena occurs. 

The preliminary validation by numerical simulation with OpenFOAM is also carried out in this article. Since the $k-\varepsilon-\zeta$ model contains 17 equations in the 2D-3C turbulence structure, which is more difficult to solve and make convergence than the second-order moment model in the aspect of solving nonlinear equations. Hence a numerical approximation of the production term in turbulent kinetic energy equation is made to enhance the stability and convergence during solving this model. We first analyze the characteristics of the $k-\varepsilon-\zeta$ model by simulating the turbulent flow in the two-dimensional straight channel. The results show that $k-\varepsilon-\zeta$ model can predict the anisotropy of Reynolds stress. The Lumley triangle has also been validated for $k-\varepsilon-\zeta$ model, and the eigenvalue distribution of the dimensionless Reynolds stress anisotropy tensor is in good agreement with DNS data. And then, we also calculate and verify the two-dimensional complex turbulent flow in a plane diffuser. The results show that $k-\varepsilon-\zeta$ model can effectively resolve the separation bubble. The detachment point and reattachment point are also quantitatively checked, which are in good agreement with the experimental measurements and LES calculation results. Compared with the standard $k-\varepsilon$ model, the $k-\varepsilon-\zeta$ model has a better analytical ability for complex turbulent flow.

Future research is needed to calibrate the damping function $f\left(Re_t, \mathbf{u}, \nabla \mathbf{u}, \cdots\right)$ with experimental measurements or DNS datasets. The form and arguments of the damping function can be predicted by neural network \cite{Maulik2017,Xie2021,Fang2020} or machine learning \cite{Fukami2019,Fukami2020,Li2020,Pandey2020}. In addition, other iterative algorithms also can be studied to enhance the stability and convergence of during solving $k-\varepsilon-\zeta$ model. When the damping function is calibrated or other more stable iterative algorithms are developed, more severe and more detailed benchmark tests of the $k-\varepsilon-\zeta$ model should be carried out, to provide more useful information on high-order eddy viscosity tensor. These important developments and tests of $k-\varepsilon-\zeta$ model will be the subject of a future study.

\appendix
\section{Tensor function derivative of source terms in Reynolds viscous stress equation.}
In Section 2.2, we carry out the derivation of the tensor function derivative of $P_{ij}^M$ in the Reynolds viscous stress transport equation to the velocity gradient tensor. With the mathematical transformation and assumption in the derivation process, we can obtain the tensor function derivative of other source terms in the Reynolds viscous stress transport equation to the velocity gradient tensor, as
\begin{subequations}
  \begin{equation}
    {\mathbf{D}^M}' = \oint_A n_m\left(C_S \frac{k^2}{\varepsilon}\right)\frac{\partial \zeta_{ijkl}}{\partial x_m} dA,
  \end{equation}
  \begin{equation}
    \begin{aligned}
      {\mathbf{\Phi}_S^M}' &= \int_V C_2\left[\left(2\delta_{ij}R_{\underline{i}m}^V \delta_{\underline{i}l}\delta_{mk} + 2\delta_{ij}\zeta_{\underline{i}mkl}\frac{\partial u_{\underline{i}}}{\partial x_m}\right) + \epsilon_{ij}R_{\underline{i}\underline{i}}^V \Omega_{kl} + \epsilon_{ij}\omega_Z\zeta_{\underline{i}\underline{i}kl}\right]dV   \\
      &-\int_V C_2 \frac{2}{3}k\left(\delta_{ik}\delta_{jl} + \delta_{il}\delta_{jk}\right)dV  \\
      &-\int_V C_2 \frac{2}{3}\delta_{ij}\left(\frac{\partial u_r}{\partial x_s}\zeta_{rskl} + R_{rs}^V\delta_{rl}\delta_{sk}\right) dV,
    \end{aligned}
  \end{equation}
  \begin{equation}
    {\mathbf{P}_k^M}' = \int_V \frac{2}{3}\delta_{ij}\left(\frac{\partial u_r}{\partial x_s} \zeta_{rskl} + R_{rs}^V\delta_{rl}\delta_{sk}\right)dV,
  \end{equation}
  \begin{equation}
    {\mathbf{D}_k^M}' = \mathbf{0}
  \end{equation}
\end{subequations}
All above derivatives are in the base of $\mathbf{e}_i\mathbf{e}_j\mathbf{e}_k\mathbf{e}_l$.

\section*{Acknowledgement(s)}
The authors are grateful to Prof. Nan Jiang at Tianjin University for the useful discussions and suggestions. 

\section*{Funding}
This work is supported by the National Key Research and Development Program of China (No. 2020YBF1902000).

\bibliographystyle{unsrt}
\bibliography{References.bib}
\end{document}